\documentclass[aps,prd,showpacs,superscriptaddress,nofootinbib,amsmath,amssymb]{revtex4}
\usepackage{graphicx}
\usepackage{dcolumn}
\usepackage{bm,latexsym}
\usepackage{mathrsfs}

\def\Journal#1#2#3#4{{#1} {\bf #2}, #3 (#4)}

\def\APJ{{Astrophys. J.}}
\def\APJS{{Astrophys. J. Suppl. Ser.}}

\def\CQG{{Class. Quant. Grav.}}

\def\JMP{{J. Math. Phys.}}
\def\LivRev{{Living Rev. Relativity}} 
\def\PLB{{Phys. Lett.}  B}
\def\PLA{{Phys. Lett.}  A}
\def\PRL{Phys. Rev. Lett.}
\def\PREV{Phys. Rev.}
\def\PRD{{Phys. Rev.} D}

\def\NAT{{Nature}}
\def\NATL{{Nature (London)}}

\def\RMF{{Rev. Mex. F\'\i s.}}

\def\be{\begin{equation}}
\def\ee{\end{equation}}
\def\bea{\begin{eqnarray}}
\def\eea{\end{eqnarray}}

\begin{document}

\title{The Cauchy problem of scalar-tensor theories of gravity}
\author{Marcelo Salgado} 
\email{marcelo@nucleares.unam.mx}
\homepage{http://www.nucleares.unam.mx/~marcelo}
\affiliation{Instituto de Ciencias Nucleares 
\\ Universidad Nacional Aut\'onoma de M\'exico 
\\ Apdo. Postal 70--543 M\'exico 04510 D.F., M\'exico}

\date{\today}

\begin{abstract}
The 3+1 formulation of scalar-tensor theories of gravity (STT) 
is obtained in the physical (Jordan) frame departing from the 4+0 
covariant field equations. Contrary to the common belief (folklore), 
the new system of ADM-like equations shows that the Cauchy problem of STT 
is well formulated (in the sense 
that the whole system of evolution equations is of first order in the time-derivative). This is 
the first step towards a full first order (in time and space) formulation 
from which a subsequent hyperbolicity analysis (a well-posedness determination) 
can be performed. Several gauge (lapse and shift) conditions 
are considered and implemented for STT. In particular, a generalization of the 
harmonic gauge for STT allows us to prove the well posedness of the STT using a 
second order analysis which is very similar to the one used in general relativity.
Some spacetimes of astrophysical and 
cosmological interest 
are considered as specific applications. Several appendices complement 
the ideas of the main part of the paper.
\end{abstract}

\maketitle

\section{Introduction}

In the last decade there has been an increasing interest in the so-called
scalar-tensor theories of gravity (STT; see Ref. \cite{damour1} for a
review) in view of the possible deviations that Einstein's general theory of
relativity (GR) could show in the framework of several upcoming
observations.  
The most prominent example of STT is perhaps the Brans-Dicke (BD) theory 
which intends to incorporate the Mach's principle into GR 
by considering a varying gravitational constant \cite{BD}.

Scalar-tensor theories are alternative theories of gravity in which 
a scalar-field is coupled non-minimally (NMC) to the curvature 
(in the Jordan-frame representation -see Sec. II-). Due to this 
fact and also due to the mere presence of a fundamental scalar field, 
these kind of theories predict several phenomena not present in GR. In particular, 
there has been a considerable interest in
detecting a scalar-wave component (spin-0 waves) in addition to the ordinary
gravitational waves (spin-2 waves) predicted by GR (see for instance 
Refs. \cite{scalwaves,SNN,HCNN}). Among the potential emitters of such
scalar-waves are the compact binary systems and neutron stars. On the 
other hand, if no appreciable deviation of GR are observed, then the 
experiments will be equally helpful to bound the 
STT parameters or couplings \cite{damour2,nagata,will1}. Through the
effects of spontaneous scalarization in neutron stars (a new phenomenon 
predicted by STT) 
\cite{damour2,damour3,novak, salgado2}, it seems that
scalar-waves have a chance to be observed in detectors like VIRGO
or LIGO within a few hundreds of kiloparsecs \cite{novak}. Moreover, there is
the hope that resonant mass detectors of spherical shape or interferometers
like LISA can resolve the existence of a scalar-gravitational wave 
\cite{scalwaves,will1}. Furthermore, scalar-waves produced during oscillations 
in neutron stars could leave an imprint on the spin-2 waves spectrum and 
therefore spin-0 waves could be indirectly detected \cite{kokkotas}.

At the large scale (cosmological scales), STT have been proposed as models
for dark energy that can replace the cosmological constant 
\cite{dark,riazuelo2,salgado3}, and also as an attempt for
explaining some other features of the galaxy distribution in our universe 
\cite{us}. 

 At smaller scales (scales of the order of
meters or kilometers) the spurious discovery of a fifth force \cite{fishbach}, renewed the
interest about the existence of new fundamental fields of meter-range \cite{pimentel}. 
At this regard, considerable effort has been put in measuring 
gravitationally such kind of interactions that can mimic a varying
gravitational ``constant'' $G$. Indeed, a short range scalar field in STT 
would induce deviations of the $1/r^2$ Newtonian law due to the appearance of 
a Yukawa type of interaction (see Appendix B). For instance, at sub-millimeter scales, 
experiments using torsion-balance instruments bound such a Yukawa type of gravitational force 
\cite{adelberger}. Other tests consist in the analysis 
of gravitational signals induced by variations on the mass of a
lake \cite{Baldi}. Such experiments probe basically fields within the range 
of meters to some kilometers. Another kind of 
experiments probes variations of the Newton's 
gravitation law at scales of two Earth's radii by measuring the 
gravitational effects on the orbit of the laser-ranged LAGEOS satellite 
\cite{lageos}. At solar-system scales, several experiments bound the deviations of STT 
from GR through the post-Newtonian parameters (see Ref. \cite{will2}). 
In particular, the Cassini probe restricts the effective 
BD parameter $\omega_{{\rm BD}}> 4\times 10^4$ \cite{cassini}. 

At scales $\sim 10$ Mpc, large-scale structure surveys constrain 
also the strength of a Yukawa type of gravitational interaction \cite{jimenez}.

The WMAP data \cite{WMAP}, can in fact be used to bound the 
variation of the gravitational constant in some class of STT \cite{nagata}.
\bigskip

In view of the above research, STT are very well motivated theories from the 
phenomenological point of view. On the other hand, it is well known that 
more fundamental theories like strings, M-theory, etc., predict in the 
low energy limit the existence 
of scalar-fields that can be coupled non-minimally to the curvature 
(see Ref. \cite{maedabook} for a review).
\bigskip

From the mathematical point of view it is important to provide a solid basis for STT. For instance, a 
meaningful physical theory is expected 
to have a well formulated and well posed Cauchy problem (initial value problem) 
\footnote{Here a distinction is made between a well formulated and a well 
posed Cauchy problem. Basically a well posed problem not only requires 
that the system of partial differential equations be 
put as a system of first order in time but also as first order in space 
and that the characteristic matrix 
associated with the corresponding full first order system satisfy 
certain requirements as well [see Eq. (\ref{PDE}) and the discussion below].}. A formidable attention 
to this issue has been given to pure GR, from the pioneer investigations of  
several mathematicians (e.g. see Refs. \cite{Choquet}) until the present 
where the task of analyzing hyperbolic formulations of GR in several gauges 
have been renewed (see Ref. \cite{Reula} for a review). Hyperbolic systems are important for numerical stability. 
This in turn is relevant for the numerical analysis of astrophysical systems 
(e.g., gravitational core collapse, black hole collisions, inspirilling binaries) 
whose effects are to be confronted with the observations.

In the case of STT, the Cauchy problem has not been intensively analyzed like GR. 
The reason is that STT when expressed in the so-called Einstein frame (cf. Sec. II), give rise to 
field equations which take a very similar form as in GR \cite{damour1}, and therefore 
the well-posedness of the Einstein's field equations (in a given gauge) could in principle 
be translated to STT. In the case of very particular STT like the one proposed by O'Hanlon \cite{ohanlon}, 
it has been shown that they can be mapped to 
higher order theories of gravity where the Cauchy problem was studied \cite{teyssandier}. 
In the Jordan frame,  the well posedness of the Cauchy problem has been analyzed for two 
particular examples of STT. One is the  BD theory, analyzed 
in Ref. \cite{cocke-cohen} using {\it normal Gaussian coordinates}. The other one 
is a STT corresponding to the conformal coupling which was studied by 
Noakes \cite{noakes} using {\it harmonic coordinates}. 

Only for the special cases mentioned above the well posedness of the Cauchy problem has been studied in the 
Jordan frame. However, in that frame, none of the analysis performed so far have envisaged the possibility of writing 
the STT in full 3+1 form in order to proof a well formulated Cauchy problem 
\footnote{In the Einstein frame the 3+1 formulation of STT is almost 
straightforward since one only requires the 3+1 formulation of GR and considers the metric there 
as the non-physical metric.}. 
One exception is perhaps the analysis by 
Sheel, Shapiro \& Teukolsky \cite{SST1,SST2}, who studied numerically the BD theory in the Jordan frame 
by using a 3+1 decomposition, but this was performed only for the spherically symmetric case with 
 a very particular gauge choice and as mentioned for a very specific STT 
(more comments about this analysis are given in Sec. V). 
\bigskip

In this paper, another alternative is given for treating the Cauchy problem of STT. 
The STT treated here are of general kind and not only a specific case where the non-minimal 
coupling (NMC) function takes a particular form. Moreover, 
the field equations of STT are written directly in the Jordan (physical frame) 
and show (contrary to the common believe) that one can obtain a 3+1 formulation 
(like the Arnowitt-Deser-Misner formulation of GR \cite{ADM,york79}) leading to first order in time evolution 
equations for the gravitational and the scalar field (i.e., a well formulated 
Cauchy problem). In fact, this is the first part of a research in which the Cauchy problem of 
STT is analyzed. In a future work, the plan is to obtain a full first-order 
(both in time and space) formulation of STT in the Jordan frame and perform 
a hyperbolicity analysis using different gauges in order to 
establish the well-posedness of the Cauchy problem (see Appendix A for the 
well-posedness using a covariant approach).

In the author's view, the reason for being interested in analyzing the Cauchy 
problem in the mathematically more involved Jordan frame, relies in that such a frame 
can provide much more physical insight than that the Einstein frame
(cf. Ref. \cite{SST2} for a similar point of view). This latter 
tends sometimes to hide situations where the (true) physical metric can 
have pathologies. Furthermore, when working in the Einstein frame, one 
can encounter examples where one is not able to recover the physical 
frame due to the presence of singularities \cite{faraoni,abramo} or 
in situations where the maps from one frame to the other 
[cf. Eqs. (\ref{map1}) and (\ref{map2})] are ill-defined
\footnote{There exists also the case where the Brans-Dicke parametrization of 
STT [cf. Eq.(\ref{BD})] together with the use of the Einstein frame can lead to ill-defined 
maps between the Jordan and Einstein frames \cite{salsudnuc}.}. 
Such undesirable situations might happen when $F(\phi)$ is not positive definite. 
Clearly, under a well 
defined and physically reasonable STT both frames should be acceptable and 
it is a matter of mathematically convenience to use one over the other (cf. Ref. \cite{flanagan}). 
In this regard, it 
is to be stressed that it is not the aim of 
this paper to feed the long debate between both frames. This has 
been the issue of countless arguments (see Refs. \cite{faraonietal,flanagan,maedabook} for a review).  
The purpose here is to provide a new alternative which gives a new point 
of departure for researchers interested in numerical analysis of STT or 
in other applications where the 3+1 formulation is better suited than the covariant approach.
\bigskip

The paper is organized as follows. In Sec. II, scalar-tensor theories of gravity for a 
single scalar field are introduced in several parametrizations (including the Einstein and 
the Jordan frame representations). Nevertheless, the rest of the paper focuses only 
in the Jordan frame. 
In Sec. III, the standard 3+1 formalism is briefly reviewed and 
applied to the field equations of Sec. II. The new 3+1 equations for STT are then 
derived. In Sec. IV, several gauge choices (lapse and shift) are proposed for a further 
numerical analysis. In Sec. V, particular examples of astrophysical and cosmological 
interest are discussed. Finally in Sec. VI, several lines of research are suggested 
as a natural continuation of this work. Various appendices complement the 
paper. Appendix A shows that the Cauchy problem of STT is well posed at least in a 
particular gauge. Appendix B re-analyzes the linear limit of STT using
a covariant and a 3+1 approach. Some aspects of the weak-field limit are 
considered. Appendix C presents the 3+1 expression of differential 
operators appearing in several gauge choices. Lastly, Appendix D, sketches a self-consistency check 
performed on the 3+1 equations of Sec. III using the Bianchi identities and therefore the conservation of the effective 
energy-momentum tensor of STT.

\section{Scalar-Tensor Theories of Gravity}

The general action for STT with a single scalar field is given by 
\begin{equation}
  \label{jordan}
S[g_{ab}, \phi, \psi] = \int \left\{ \frac{1}{16\pi G_0} F(\phi) R
-\left( \frac{1}{2}(\nabla \phi)^2 + V(\phi) \right) \right\} \sqrt{-g} d^4x + S_{\rm matt}[g_{ab}, \psi] \,\,\,\,,
\end{equation}
where $\psi$ represents collectively the matter fields (fields other than $%
\phi$; units where $c=1$ are employed).

The representation of STT given by Eq. (\ref{jordan})
is called the {\it Jordan frame} representation. One can parametrize the
same theories as 
\begin{equation}
  \label{BD}
S[g_{ab}, \Phi, \psi] = \frac{1}{16\pi G_0} \int \left\{ \Phi R - \frac{%
\omega_{{\rm BD}} (\Phi)}{\Phi} (\nabla \Phi)^2 + 2\Phi \lambda (\Phi) \right\}
\sqrt{-g} d^4 x + S_{\rm matt}[g_{ab}, \psi] \,\,\,\,,
\end{equation}
where 
\begin{eqnarray}  
\Phi &:=& F(\phi) \,\,\,\,, \\
\label{BDpar}
\omega_{{\rm BD}} (\Phi) &:=& \frac{8\pi G_0\Phi}{(F^\prime)^2}
\,\,\,\,, \\
\lambda (\Phi) &:=& -\frac{8\pi G_0 V(\phi)}{\Phi} \,\,\,\,,
\end{eqnarray}
here $^\prime$ indicates $\partial_\phi$.
For instance, the Brans-Dicke theory with 
$\omega=$const. corresponds to $F= 2\pi G_0 \phi^2/\omega$ and $V(\phi) =0$.
It is also customary to parametrize STT in the
so-called {\it Einstein frame} by introducing {\it non-physical} fields as
follows, 
\begin{eqnarray}
\label{map1}
g_{ab}^* &:=& F(\phi) g_{ab} \,\,\,\,, \\
\label{map2}
\phi^* &:=& \int \left[ \frac{3}{4}\left(\frac{F^\prime}{F}\right)^2 + 
\frac{4\pi G_0}{F(\phi)}\right]^{1/2} d\phi \,\,\,\,, \\
W(\phi^*)&:=& \frac{4\pi G_0 V^*(\phi^*)}{{F^*}^2} \,\,\,\,, \\
F^*(\phi^*) &=& F(\phi) \,\,\,\,,
\end{eqnarray}
so that the action Eq. (\ref{jordan}) takes the form 
\begin{equation}
 \label{einst}
S[g^*_{ab}, \phi^*, \psi] = \frac{1}{16\pi G_0} \int \left[\rule{0mm}{0.5cm} R^* - 2
(\nabla^* \phi^*)^2 - 4W(\phi^*) \right]\,\sqrt{-g^*} d^4 x + 
S_{\rm matt}[g_{ab}^*/F^*(\phi^*), \psi] \,\,\,\,,
\end{equation}
where all quantities with `*' are computed with the non-physical metric $%
g_{ab}^*$ and $\phi^*$.

One can remark that although the field equations obtained from the
Einstein frame are simpler than those from the Jordan frame, in the sense
that the field $\phi ^{*}$ appears to be coupled minimally to the
non-physical metric, the matter equations derived from the Bianchi
identities $\nabla _{c }^{*} G^{*\,c a }=0$ will have sources, i.e., $%
\nabla _{c }^{*}T_{\psi }^{*\,\,c a }\neq 0$, where $T_{\psi
}^{*\,\,a b }=T_{\psi }^{a b }/{F^{*}}^{3}(\phi ^{*})$ is the non
physical energy-momentum tensor of the matter fields $\psi $. Nevertheless, in
the Jordan frame the matter equations resulting from the Bianchi identities $%
\nabla _{c }G^{c a }=0$ turn to satisfy $\nabla _{c }T_{\psi }^{ca}=0$, reflecting explicitly the fulfillment of the Einstein's weak 
equivalence principle (this is the origin of the name ``physical metric'').

In the following only the Jordan frame representation of STT 
will be used (for a detailed analysis of STT in the Einstein 
frame see Ref. \cite{damour1}). The field equations obtained from 
the action (\ref{jordan}) are\footnote{Latin indices from the first 
letters of the alphabet $a,b,c,...$ are four-dimensional and run $0-3$. 
Latin indices starting from $i$ $(i,j,k,...)$ are three-dimensional and 
run $1-3$.} 
\begin{eqnarray}  
\label{Einst}
G_{ab} &=& 8\pi G_0 T_{ab}\,\,\,\,, \\
\label{KGo}
\Box \phi &+& \frac{1}{2}f^\prime R = V^\prime \,\,\,,
\end{eqnarray}
where $G_{ab}= R_{ab}-\frac{1}{2}g_{ab}R$ and
\begin{eqnarray}  
\label{effTmunu}
T_{ab} &:=& \frac{G_{{\rm eff}}}{G_0}\left(\rule{0mm}{0.5cm} T_{ab}^f + 
T_{ab}^{\phi} + T_{ab}^{{\rm matt}}\right)\,\,\,\,, \\
\label{TabF}
T_{ab}^f &:= & \nabla_a\left(f^\prime 
\nabla_b\phi\right) - g_{ab}\nabla_c \left(f^\prime 
\nabla^c \phi\right) \,\,\,\,, \\
T_{ab}^{\phi} &:= & (\nabla_a \phi)(\nabla_b \phi) - g_{ab}
\left[ \frac{1}{2}(\nabla \phi)^2 + V(\phi)\right ] \,\,\,\,, \\
\label{Geff}
G_{{\rm eff}} &:=& \frac{1}{8\pi f} \,\,\,\,,\,\,\,\,f:=\frac{F}{8\pi G_0} 
\,\,\,\,.
\end{eqnarray}
Using Eq. (\ref{Einst}), the Ricci scalar can be expressed in terms of the 
energy momentum tensor Eq. (\ref{effTmunu}) and then Eq.(\ref{KGo}) takes the following form,
\begin{widetext}
\begin{equation}
\label{KG}
{\Box \phi} = \frac{ f V^\prime - 2f^\prime V -\frac{1}{2}f^\prime
\left( 1 +  3f^{\prime\prime} 
\right)(\nabla \phi)^2 + \frac{1}{2}f^\prime T_{{\rm matt}} }
{f\left(1 + \frac{3{f^\prime}^2}{2f}\right) }\,\,\,\,,
\end{equation}
\end{widetext}
where $T_{{\rm matt}}$ stands for the trace of $T^{ab}_{{\rm matt}}$ and
the subscript ``matt'' refers to the matter fields (fields other that $\phi$).

Now, the Bianchi identities imply 
\begin{equation}
\nabla _{c }T^{c a }=0\,\,\,\,.
\end{equation}
However, the use of the field equations leads, as mentioned before, to the 
conservation equations for the matter alone 
\begin{equation}
\nabla _{c }T_{{\rm matt}}^{c a }=0\,\,\,\,,
\end{equation}
which implies in the case of test particles, that neutral bodies are subject to no
other long range forces than the gravitational ones (free falling
particles). In other words, test particles follow the geodesics of the physical metric 
and the whole effect of the scalar-field is reflected in the 
modification of the geometry. 

In the next section the field equations will be recasted in a 3+1 form 
which is specially suited for numerical applications.

\section{The 3+1 formulation}
The aim of this section is to 
consider the 3+1 or Arnowitt-Deser-Misner (ADM \cite{ADM}) formulation of general 
relativity {\it \`a la} York \cite{york79} and apply a similar formalism to reformulate the 
scalar tensor theories of gravity in a way as to obtain a set of first order time 
evolution equations and a set of constraint equations for the initial data.

One then considers 
a spacetime $(M,g_{ab})$ (assumed to be globally hyperbolic) which is 
to be foliated by a family of spacelike hypersurfaces $\Sigma_t$ parametrized by 
a global time function $t$. The foliation 
is achieved in the following way. On a Cauchy surface $\Sigma_t$, one is given 
an initial data set that satisfy some constraint equations, and then the spacetime is 
``reconstructed'' by evolving the initial data 
using a suitable set of evolution equations. This set of constraints and 
evolution equations are known as the ADM equations of GR. In order to find the 
equivalent set of equations for the case of STT a similar algebraic and geometric 
decomposition of the field equations (\ref{Einst}) and (\ref{KG}) 
have to be performed. In order to do so, 
some basic aspects of the usual 3+1 decomposition of the Einstein equations are introduced.

The procedure for obtaining the 
3+1 splitting of the Einstein equations consists in projecting 
tensor fields in the direction parallel and orthogonal to the timelike unit vector field 
$n^a$ ($n^a n_a=-1$) which is normal to $\Sigma_t$. The projection onto $\Sigma_t$ is performed 
by defining the {\it projector} 
\begin{equation}
h^a_{\,\,b} = \delta^a_{\,\,b} + n^a n_b \,\,\,\,.
\end{equation}
The property of this tensor field is that it is idempotent $h_a^{\,\,c}h_c^{\,\,b}=
 h_a^{\,\,b}$. A tensor field $\,\!^3T^{a_1 a_2 ... a_k}_{\hspace{1.1cm}b_1 b_2...b_l}$ is said to 
be tangent to $\Sigma_t$ if contracted with $n^a$ is zero or if contracted 
with $h^a_{\,\,b}$ remains unchanged. For brevity, this tensors will be termed as 
$3-$tensors. Any tensor field can be decomposed {\it orthogonally} 
using $h^a_{\,\,b}$ and $n^a$. In particular, a 4-vector
$w^a \in T_p^M$ is decomposed as follows 
\begin{equation}
w^a =\,\!^3w^a + w_\perp n^a \,\,\,\,,
\end{equation}
where $\,\!^3w^a:= h^a_{\,\,b} w^b$ and $w_\perp:= -n_c w^c$, with 
${\rm sign} g_{ab}= (-\,,\,+\,,\,+\,,\,+)$. Moreover, the 
3+1 splitting of the metric is as follows:
\begin{equation}  \label{3+1metric}
ds^2= -(N^2 - N^iN_i)dt^2 - 2N_i dt dx^i + h_{ij} dx^i dx^j\,\,\,,
\end{equation}
where the {\it lapse} function $N>0$ is defined as to normalize the (future pointing) 
dual vector field $n_a= -N\nabla_a t$. The shift vector is given by 
$N^a:= -h^a_{\,\,b} t^b $, where 
$t^a$ is a vector field that represents the ``flow'' of the time lines and that 
satisfies $t^a\nabla_a t=1$. This means that 
$t^a$ is orthogonally decomposed as $t^a= -N^a + Nn^a$, with $N=-n_a t^a$. 
Here $h_{ij}$ is the 3-metric (or induced metric) of the manifold $\Sigma_t$.

Another important object is the extrinsic curvature of the embedings 
$\Sigma_t$ which is defined as
\begin{equation} 
\label{K_ab}
K_{ab}:= -\frac{1}{2} {\cal L}_{\mbox{\boldmath{$n$}}}h_{ab} \,\,\,\,,
\end{equation} 
where ${\cal L}_{\mbox{\boldmath{$n$}}}$ stands for the Lie derivative along 
$n^a$. From the above definition one can obtain the following identity
\begin{equation}
K_{ab}= - h^{\,\,c}_{a}  h^{\,\,d}_{b} \nabla_c n_d \,\,\,,
\end{equation}
which shows that $K_{ab}$ is a 3-tensor field.
 
As it is well known, the set $(\Sigma_t, h_{ab},K_{ab})$ provides the 
initial data for the gravitational field. This data cannot be arbitrary but 
have to satisfy the Einstein constraint equations (see below).

At this point it is useful to 
introduce a derivative operator compatible with $h_{ab}$. Given 
a 3-tensor field $\,\!^3T^{a_1 a_2 ...a_k}_{\hspace{1.1cm}b_1 b_2 ... b_l}$, one defines 
\begin{equation}
D_e \,\!^3T^{a_1 a_2 ... a_k}_{\hspace{1.1cm} b_1 b_2 ... b_l}= 
h^{a_1}_{\,\,c_1} h^{a_2} _{\,\,c_2}... h^{a_k} _{\,\,c_k} 
h^{\,\,d_1}_{b_1} h^{\,\,d_2}_{b_2}
...h^{\,\,d_l}_{b_l} h^{f}_{e} \nabla_f \,\!^3T^{c_1 c_2...c_k}_{\hspace{1.1cm}d_1 d_2 ... d_l}\,\,\,,
\end{equation}
where $D_c h_{cd} \equiv 0$. Finally, it is to be mentioned that the indices of 
3-tensors can be raised and lowered with $h^{ab}$ and $h_{ab}$. 

For the purpose of this paper, it is convenient to introduce the following quantities 
\begin{eqnarray}
\label{Q}
Q_a &:=& D_a \phi \,\,\,,\\
\Pi &:=& {\cal L}_{\mbox{\boldmath{$n$}}} \phi \,\,\,,
\label{Pidef}
\end{eqnarray}
where $\Pi$ is to be identified as the ${\it momentum}$ of the scalar field which 
plays a similar role as $K_{ab}$ for the gravitational field. On the other 
hand $Q_a$ is the analogous of the 3-Christoffel symbols in the sense that it contains first order spatial 
derivatives of the scalar-field. Nonetheless, unlike the 3-Christoffel symbols, 
$Q^a$ is as a 3-vector field.

Now from Eqs. (\ref{K_ab}) and (\ref{Pidef}) one obtains the following useful expressions
\begin{eqnarray}  
\label{K_ij}
K_{ij} &=& -\nabla_i n_j= -N\Gamma^t_{ij} =-\frac{1}{2N}\left( \frac{\partial
h_{ij}}{\partial t} + \,D_j N_i + \,D_i N_j \right) \,\,\,,\\
\Pi &=& n^a\nabla_a\phi = \frac{1}{N}\left(\rule{0mm}{0.5cm} \partial_t \phi + N^a Q_a\right) \,\,\,.
\label{Pi}
\end{eqnarray}
Eq. (\ref{K_ij}) is to be regarded as an evolution equation for 
$h_{ij}$, while (\ref{Pi}) provides the evolution equation for 
$\phi$. In fact, one can obtain an evolution equation for $Q_i$ by applying 
$D_i$ on ({\ref{Pi}) and using $\partial_{it}^2 \phi = \partial_{ti}^2 \phi= 
\partial_t Q_i$, where in this case, $D_i= h^{c} _{i} \partial_c 
\equiv \partial_i$. Therefore 

\begin{equation}
\label{evQ1}
\partial_t Q_i = D_i (N\Pi) - D_i (N^l Q_l) \,\,\,.
\end{equation}
Because of the symmetry $D_i Q_j= D_j Q_i$ 
which follows from the integrability condition $\partial_{ij}^2\phi = \partial_{ji}^2 \phi$, one 
can write
\begin{equation}
\label{evQ2}
\partial_t Q_i + N^l\partial_l Q_i + Q_l\partial_i N^l = D_i (N\Pi)  \,\,\,.
\end{equation}
This is indeed equivalent to  
\begin{equation}
\label{evQ3}
 {\cal L}_{\mbox{\boldmath{$n$}}} Q_a = \frac{1}{N}D_a(N\Pi) \,\,\,.
\end{equation}

The orthogonal decomposition of the energy-momentum tensor is given by \cite{york79}:

\begin{equation}  
\label{Tort}
T^{ab} = S^{ab} + J^{a} n^{b} + n^a J^{b} + E n^a n^b\,\,\,\,,
\end{equation}
where
\bea \label{S}
S^{ab }&:=&\,h_{\,\,\,c}^a h_{\,\,\,d}^b T^{cd}
\,\,\,\,\,, \\
\label{J}
J^{a}&:=&  - h_{\,\,\,c}^a  T^{c d} n_d 
 \,\,\,\,\,\,\,\,\,\,\,\,\,\,\,\,,\\
\label{E}
E &:=& T^{cd} n_c n_d 
\,\,\,\,\,\,\,\,\,,
\eea
here $S^{ab}$ is the 3-energy-momentum tensor,
 $J^a$ is the 3-{\it momentum density vector} and $E$ is the 
{\it total energy density} measured by the observer 
orthogonal to $\Sigma_t$ with 4-velocity $n^a$. 

In this paper $T^{ab}$
is the effective energy-momentum tensor (\ref{effTmunu})
 which includes the contribution of the scalar field and another kind of matter: 
$T^{ab}= \sum_i T^{ab}_i $. This means that 
\begin{equation}  
\label{3+1mattvartot}
E = \sum_i E_i \,\,\,,\,\,\, J^a = \sum_i J^a _i \,\,\,,\,\,\,
S^{ab} = \sum_i S^{ab}_i \,\,\,\,.
\end{equation}
Specifically for the analysis at hand the above quantities are given by
\begin{eqnarray}
\label{ESTT}
E &=& \frac{G_{\rm eff}}{G_0}\left(\rule{0mm}{0.4cm} E^f + E^\phi + E^{\rm matt}\right)\,\,\,,\\
J_a &=& \frac{G_{\rm eff}}{G_0}\left( \rule{0mm}{0.4cm} J^f_a + J^\phi_a + 
J^{\rm matt}_a\right)\,\,\,,\\
S_{ab} &=& \frac{G_{\rm eff}}{G_0}\left( \rule{0mm}{0.4cm} S^f_{ab} + S^\phi_{ab} + 
S^{\rm matt}_{ab}\right)\,\,\,. \label{SabSTT}
\end{eqnarray}

In order to proceed with the 3+1 formulation of STT, one is required to 
project orthogonally the Einstein equations (\ref{Einst}).  
The projection of Einstein equations $G_{ab}= 8\pi G_0 T_{ab}$ 
[or  $R_{ab}= 4\pi G_0\left(2T_{ab}-
T^{c}_{\,\,\,c} g_{ab}\right)$, better suited to obtain the dynamic equations] 
in the directions tangent and
orthogonal to $\Sigma_t$, followed by the use of the Gauss-Codazzi 
equations leads to the 3+1 form of Einstein equations {\it \`a la} York \cite{york79}:

\begin{equation}  \label{CEHf}
^3 R + K^2 - K_{ij} K^{ij}= 16\pi G_0 E \,\,\,,
\end{equation}
known as the Hamiltonian constraint.

\begin{equation}  \label{CEMf}
D_l K_{\,\,\,\,\,i}^{l} - D_i K= 8\pi G_0 J_i \,\,\,\,,
\end{equation}
known as the {\it momentum constraints}.

The {\it dynamic} Einstein equations read\footnote{This can be written as 
\begin{equation}
h_{\,\,\,c}^a h^{\,\,\,d}_b {\cal L}_{\mbox{\boldmath{$n$}}} K_{\,\,\,d}^c
+ \frac{1}{N}\,D^a\,D_b N -\,^3 R_{\,\,\,b}^a - K K_{\,\,\,b}^a  
= 4 \pi G_0 \left[\rule{0mm}{0.4cm} (S-E)h^a_{\,\,b} - 2S^a_{\,\,b}\right]\,\,\,\,,
\end{equation}
or alternatively as
\begin{equation}
 {\cal L}_{\mbox{\boldmath{$n$}}} K_{ab}  
 + \frac{1}{N}\,D_a\,D_b N  -\,^3 R_{ab}  -  K K_{ab} + 2 K_{ac}K_{\,\,\,b}^c  
 = 4 \pi G_0 \left[\rule{0mm}{0.4cm} (S-E)h_{ab} - 2S_{ab}\right]\,\,\,\,,
\end{equation}
where ${\cal L}_{\mbox{\boldmath{$n$}}}K_{ab}= \frac{1}{N}\left({\cal L}_{\mbox{\boldmath{$t$}}}K_{ab} + 
{\cal L}_{\mbox{\boldmath{$N$}}}K_{ab}\right)$ or 
$h_{\,\,\,c}^a h^{\,\,\,d}_b {\cal L}_{\mbox{\boldmath{$n$}}} K_{\,\,\,d}^c
= \frac{1}{N}h_{\,\,\,c}^a h^{\,\,\,d}_b\left({\cal L}_{\mbox{\boldmath{$t$}}}K_{\,\,\,d}^c + 
{\cal L}_{\mbox{\boldmath{$N$}}}K_{\,\,\,d}^c\right)$}
\begin{eqnarray}  
\label{EDEf}
& & \partial_t K_{\,\,\,j}^i + N^l \partial_l K_{\,\,\,j}^i + K_{\,\,\,l}^i
\partial_j N^l - K_{\,\,\,j}^l \partial_l N^i \nonumber 
 + \,D^i\,D_j N  -\,^3 R_{\,\,\,j}^i N - N K K_{\,\,\,j}^i  \nonumber \\
&& = 4 \pi G_0 N\left[\rule{0mm}{0.4cm} (S-E)\delta^i_{\,\,j} - 2S^i_{\,\,j}\right]\,\,\,\,,
\end{eqnarray}
where $K$ and $S$ stand for the traces of the extrinsic curvature and the 
3-energy-momentum tensor respectively. All
the quantities written with a `$3$' index refer to those computed with the
three-metric $h_{ij}$. 

An evolution equation for the trace $K$ is obtained by taking the trace in
Eq. (\ref{EDEf}) followed by the use of Eq. (\ref{CEHf}) to give 
\begin{equation}  \label{EDK2}
\partial_t K + N^l \partial_l K + \,^3\Delta N -N K_{ij} K^{ij} = 4 \pi G_0
N\left[\rule{0mm}{0.4cm} S + E\right] \,\,\,,
\end{equation}
where $\,^3\Delta:= D^l D_l$ stands for the Laplacian operator compatible with $h_{ij}$.
\bigskip

As mentioned at the beginning of this section, in pure and vacuum 
GR, given an initial data set $(\Sigma_t, h_{ij}, K_{ij}) $ satisfying the constraints
Eqs. (\ref{CEHf}) and (\ref{CEMf}), one performs the Cauchy development by 
evolving $h_{ij}$ and $K_{ij}$ forward in time using Eqs. (\ref{K_ij}) and 
(\ref{EDEf}), respectively. Of course, in order to achieve the evolution, lapse and shift 
conditions (gauge conditions) have to be imposed (see Sec. IV).
\bigskip

Now, although the Einstein equations give in general no information 
about the specific form of the energy-momentum tensor (EMT), 
one usually considers some features on $T_{ab}$ that leads to a well posed 
initial value problem. In particular is common to consider EMT that 
does not contain second order time derivatives (SOTD) of the matter fields 
so that the theorems for existence and uniqueness of solutions of 
Einstein equations together with the matter 
equations $\nabla_a T^{ab}=0$ can be applied (e.g. see Refs. \cite{hawkellis,wald} ). 

Now, in the case of STT, the non-minimal coupling (NMC) of the scalar-field with the 
curvature makes however that the right-hand-side (r.h.s) of Eq. (\ref{Einst}) 
actually contain second order derivatives of $\phi$ both in time and space. 
Therefore, a well formulated initial value problem seems in jeopardy.
In order 
to obtain a nice initial value formulation, such terms (notably, the SOTD terms) 
should be eliminated using the remaining field equations. 
In what follows such elimination is performed [namely, in Eq. (\ref{TabF})] 
in order to obtain the explicit form of the 3+1-effective-EMT. The remaining contributions of the EMT 
Eq. (\ref{effTmunu}) are quite standard and so their 3+1 decomposition pose no big
problem. In this way the goal is to isolate the pure matter terms 
and the scalar-field sources at the r.h.s of 
Eqs. (\ref{CEHf})$-$(\ref{EDEf}), and then 
to place the other quantities associated with the scalar field in the 
left-hand-side (l.h.s.) so that the scalar 
field be treated more less at the same footing as the gravitational field.

So let us proceed to compute $E^f= n^a n^b T_{ab}^f$. 
A straightforward but lengthy calculation leads to 
\begin{equation}
\label{EF}
E^f=  f^\prime \left(\rule{0mm}{0.4cm} \,\!^3\Delta\phi  + 
K n^a \nabla_a\phi\right) +  f^{\prime\prime} (D \phi)^2 \,\,\,\,,
\end{equation}
where $(D \phi)^2:= (D^c\phi)(D_c\phi)$. Note that $E^f$ no longer contains SOTD.

Now one computes $S_f^{ab}= h^a_{\,\,c} h^b_{\,\,d} T^{cd}_f$. An equally lengthy 
computation yields
\begin{equation}
S_f^{ab}= f^\prime  \left\{\rule{0mm}{0.4cm} D^a\,D^b\phi + 
K^{ab} (n^c \nabla_c\phi) - h^{ab}\Box\phi\right\} 
- f^{\prime\prime} \left\{ 
 h^{ab} \left[\rule{0mm}{0.4cm} (D \phi)^2- 
(n^c \nabla_c\phi)^2 \right] - (D^a \phi)(D^b \phi)\right\}\,\,\,.
\label{SF}
\end{equation}
As one can appreciate, SOTD appear in $S_F^{ab}$ only through $\Box\phi$. However, by using the r.h.s of Eq. (\ref{KG}) one 
can replace the d'Alambertian in terms of first-order derivatives and sources. 
This is done below. On the other hand, 
the term $D^aD^b\phi$ contains only second order spatial derivatives (SOSD).

From the above expression one can easily calculate its trace
\begin{equation}
\label{TrSF}
S^f= f^\prime  \left[ \rule{0mm}{0.4cm} \,\!^3\Delta\phi + 
K (n^c \nabla_c\phi) - 3 \Box\phi\right] 
+ f^{\prime\prime}  \left[\rule{0mm}{0.4cm} 
3(n^a \nabla_a\phi)^2 -2(D \phi)^2 \right] \,\,\,.
\end{equation}

From (\ref{EF}) and (\ref{TrSF}) one can obtain the quantity
\begin{equation}
\label{S-E}
S^f - E^f= - 3 f^\prime \Box\phi
-3 f^{\prime\prime}\left[\rule{0mm}{0.4cm} (D \phi)^2- (n^a \nabla_a\phi)^2\right]
\,\,\,.
\end{equation}

Finally, one requires $J_a^f= -n^d h_a^{\,\,c} T_{cd}^f$. This quantity reads
\begin{equation}
\label{JF}
J_a^f = 
- f^\prime\left[\rule{0mm}{0.4cm} K_{a}^{\,\,c}\,D_c\phi + \,D_a(n^c\nabla_c\phi)
\right] - f^{\prime\prime} (n^c \nabla_c \phi)\,D_a\phi  \,\,\,.
\end{equation}

Using Eqs. (\ref{Q}) and (\ref{Pi}) in 
Eqs. (\ref{EF})$-$(\ref{JF}) one obtains respectively 
\begin{eqnarray}
\label{EFp}
E^f &=& f^\prime \left( \rule{0mm}{0.4cm} D_a Q^a + 
K \Pi \right) +  f^{\prime\prime} Q^2  \,\,\,,\\
S_f^{ab}&=& 
f^\prime \left(\rule{0mm}{0.4cm} D^a Q^b + 
\Pi K^{ab} - h^{ab}\Box\phi\right) 
 - f^{\prime\prime} \left[\rule{0mm}{0.4cm} 
 h^{ab} \left( Q^2- \Pi^2 \right) - Q^a Q^b \right] 
\label{SFp}\,\,\,,\\
\label{TrSFp}
S^f &=& 
f^\prime\left(\rule{0mm}{0.4cm} D_a Q^a + 
K\Pi - 3 \Box\phi\right) 
+ f^{\prime\prime} \left(\rule{0mm}{0.4cm} 3\Pi ^2 -2Q^2 \right) \,\,\,,\\
\label{S-Ep}
S^f - E^f &=& -3 f^\prime \Box\phi
-3 f^{\prime\prime}\left(\rule{0mm}{0.4cm} Q^2- \Pi^2\right)
\,\,\,,\\
\label{JF2}
J_a^f &=& - f^\prime\left(\rule{0mm}{0.4cm} K_{a}^{\,\,c} Q_c + \,D_a\Pi
\right) - f^{\prime\prime} \Pi Q_a \,\,\,,
\end{eqnarray}
where $Q^2:= Q^c Q_c$.

In a similar and straightforward way one finds
\begin{eqnarray}
\label{Efip}
E_\phi &=& \frac{1}{2}\left(\rule{0mm}{0.4cm} \Pi^2 + Q^2 \right) + V(\phi) \,\,\,,\\
S_\phi^{ab}&=& Q^a Q^b -  h^{ab}\left[ \frac{1}{2}\left( Q^2- \Pi^2 \right) 
+ V(\phi) \right] 
\label{Sfip}\,\,\,,\\
\label{TrSfip}
S_\phi &=& \frac{1}{2}\left(\rule{0mm}{0.4cm} 3\Pi^2 -Q^2 \right) -3 V(\phi) \,\,\,,\\
\label{S-Efip}
S_\phi - E_\phi &=& \Pi^2 -Q^2 - 4 V(\phi)
\,\,\,,\\
\label{Jfi}
J_a^\phi &=& -\Pi Q_a \,\,\,.
\end{eqnarray}
Now, collecting these results, and 
according to Eqs. (\ref{ESTT})$-$(\ref{SabSTT}) and using Eqs. (\ref{Geff}) and 
(\ref{KG}), the total contributions of the 3+1 effective energy-momentum tensor read
\begin{eqnarray}
\label{totE}
E &=& \frac{1}{8\pi G_0 f}
\left[f^\prime \left(\rule{0mm}{0.4cm} D_c Q^c + 
K \Pi \right) + \frac{\Pi^2}{2} + \frac{Q^2}{2}
\left(\rule{0mm}{0.4cm} 1 + 2f^{\prime\prime}\right) + V(\phi) + E_{\rm matt}  \right]\,\,\,,\\
J_a &=& \frac{1}{8\pi G_0 f}\left[\rule{0mm}{0.5cm} 
-f^\prime\left(\rule{0mm}{0.4cm} K_{a}^{\,\,c} Q_c + \,D_a\Pi
\right) - \Pi Q_a\left(\rule{0mm}{0.4cm} 1 + f^{\prime\prime}\right) + 
J^{\rm matt}_a\right]\,\,\,,\\
\label{Sab}
S_{ab} &=& \frac{1}{8\pi G_0 f}\left\{\rule{0mm}{0.7cm}
Q_a Q_b\left(\rule{0mm}{0.4cm} 1+ f^{\prime\prime}\right)  
+  f^\prime \left(\rule{0mm}{0.4cm} D_a Q_b + \Pi K_{ab} \right) 
- \frac{h_{ab}}{\left(1 + \frac{3{f^\prime}^2}{2f}\right)}
\left[ \frac{1}{2}\left(\rule{0mm}{0.4cm} Q^2-\Pi^2\right)\left(1+ \frac{{f^\prime}^2}{2f}+
2 f^{\prime\prime}\right) \right. \right. \nonumber \\
&& \left. \left.
+ V\left(1- \frac{{f^\prime}^2}{2f}\right) + f^\prime V^\prime + 
\frac{{f^\prime}^2}{2f}\left(\rule{0mm}{0.4cm} S_{\rm matt}- E_{\rm matt}\right)
\right] + S^{\rm matt}_{ab} \rule{0mm}{0.7cm}\right\}\,\,\,.
\end{eqnarray}
It turns useful to write explicitly the following quantities because they appear in 
different applications
\begin{eqnarray}
\label{S-Esst}
S-E &=& \frac{1}{8\pi G_0 f \left(1 + \frac{3{f^\prime}^2}{2f}\right)}
\left[\rule{0mm}{0.5cm} \left(\rule{0mm}{0.4cm} \Pi^2 - Q^2\right)
\left(\rule{0mm}{0.4cm} 1 + 3f^{\prime\prime}\right) - 3f^\prime V^\prime -4 V(\phi) 
+ S_{\rm matt}- E_{\rm matt}  \right]\,\,\,,\\
\label{S+E}
S+E &=& \frac{1}{8\pi G_0 f}\left\{\rule{0mm}{0.7cm}  2f^\prime \left(\rule{0mm}{0.4cm} D_c Q^c + 
\Pi K \right) 
+ \frac{1}{\left(1 + \frac{3{f^\prime}^2}{2f}\right)}
\left[ 2\Pi^2\left( 1 + \frac{3{f^\prime}^2}{4f} +\frac{3f^{\prime\prime}}{2}\right)
+ 2Q^2\left(\frac{3{f^\prime}^2}{4f}\left(1 + 2f^{\prime\prime}\right)
 -\frac{f^{\prime\prime}}{2}\right)
 \right. \right. \nonumber \\
&& \left. \left.
- 2V\left(1- \frac{3{f^\prime}^2}{2f}\right) - 3 f^\prime V^\prime + 
S_{\rm matt}+ E_{\rm matt}\left(1 + \frac{3{f^\prime}^2}{f} \right)
\,\, \right] \,\, \rule{0mm}{0.7cm}\right\}\,\,\,.
\end{eqnarray}

The input of these expressions in Eqs. 
(\ref{CEHf}), (\ref{CEMf}), (\ref{EDEf}) yields finally the 3+1 equations of 
the scalar-tensor theories of gravity given by the action (\ref{jordan}). 
The Hamiltonian constraint, 
the momentum constraints and the dynamic Einstein equations of STT are 
respectively,
\begin{equation}  
\label{CEHfSST}
^3 R + K^2 - K_{ij} K^{ij} 
- \frac{2}{f}\left[
 f^\prime \left(\rule{0mm}{0.4cm} D_l Q^l + 
K \Pi \right) + \frac{\Pi^2}{2} + \frac{Q^2}{2}
\left(\rule{0mm}{0.4cm} 1 + 2f^{\prime\prime}\right) \right]
  = \frac{2}{f} \left[\rule{0mm}{0.4cm} E_{\rm matt} + V(\phi)\right]  \,\,\,,
\end{equation}

\begin{equation}  
\label{CEMfSST}
 D_l K_{\,\,\,\,\,i}^{l} - D_i K
+\frac{1}{f}\left[\rule{0mm}{0.5cm}
f^\prime\left(\rule{0mm}{0.4cm} K_{i}^{\,\,l} Q_l + \,D_i\Pi
\right) + \Pi Q_i\left(\rule{0mm}{0.4cm} 1 + f^{\prime\prime}\right)\right]
= \frac{1}{f} J_i^{\rm matt} \,\,\,\,,
\end{equation}

\begin{eqnarray}  
\label{EDEfSST}
& & \partial_t K_{\,\,\,j}^i + N^l \partial_l K_{\,\,\,j}^i + K_{\,\,\,l}^i
\partial_j N^l - K_{\,\,\,j}^l \partial_l N^i 
+ D^i D_j N  -\,^3 R_{\,\,\,j}^i N - N K K_{\,\,\,j}^i
\nonumber \\ 
&&  
+ \frac{N}{f}\left[\rule{0mm}{0.6cm} Q^i Q_j\left(\rule{0mm}{0.4cm} 1+ f^{\prime\prime}\right)  
+  f^\prime \left(\rule{0mm}{0.4cm} D^i Q_j + \Pi K^{i}_{\,\,\,j} \right) \right]  
- \frac{\delta_{\,\,\,j}^i N}{2f\left(1 + \frac{3{f^\prime}^2}{2f}\right)}
\left(\rule{0mm}{0.6cm} Q^2-\Pi^2\right)\left( \frac{{f^\prime}^2}{2f} - f^{\prime\prime}\right) 
\nonumber \\
&& = -\frac{N}{2f\left(1 + \frac{3{f^\prime}^2}{2f}\right)}
\left\{ 2 S_{{\rm matt}\,\,\,j}^i \left(1 + \frac{3{f^\prime}^2}{2f} \right)
+ \delta_{\,\,\,j}^i \left[ f^\prime V^\prime + 2V\left(1 + \frac{{f^\prime}^2}{2f} 
\right) - \left(\rule{0mm}{0.4cm} S_{\rm matt}- E_{\rm matt}\right) \left(1 + \frac{{f^\prime}^2}{f} \right)
\right]\right\}\,.
\end{eqnarray}

On the other hand, Eq. (\ref{EDK2}) reads
\begin{eqnarray}  
\label{EDKSTT}
&& \partial_t K + N^l \partial_l K + \,^3\Delta N -N K_{ij} K^{ij} 
- \frac{Nf^\prime}{f}  \left(\rule{0mm}{0.4cm} D_l Q^l + \Pi K \right) \nonumber \\
&& 
-\frac{N}{f\left(1 + \frac{3{f^\prime}^2}{2f}\right)}
\left\{ \Pi^2 \left( 1 + \frac{3{f^\prime}^2}{4f} + \frac{3f^{\prime\prime}}{2} \right)
+ Q^2\left[ \frac{3{f^\prime}^2}{4f}\left(\rule{0mm}{0.4cm} 1 + 2f^{\prime\prime}\right)  
-\frac{f^{\prime\prime}}{2}\right]  \right\} \nonumber \\
&& = \frac{N}{2f\left(1 + \frac{3{f^\prime}^2}{2f}\right)}
\left\{ S_{\rm matt} + E_{\rm matt} \left(1 + \frac{3{f^\prime}^2}{f}\right) 
- 2V\left(1- \frac{3{f^\prime}^2}{2f}\right) - 3f^\prime V^\prime \right\}
 \,\,\,.
\end{eqnarray}
\bigskip

Finally, Eq. (\ref{KG}) takes the form (c.f. Appendix C)
\begin{equation}
\label{evPi1}
{\cal L}_{\mbox{\boldmath{$n$}}} \Pi  - \Pi K - Q^c D_c[{\rm ln}N]  - D_c Q^c  
= - \frac{ f V^\prime - 2f^\prime V -\frac{1}{2}f^\prime
\left( 1 +  3f^{\prime\prime} 
\right)\left(\rule{0mm}{0.4cm} Q^2-\Pi^2\right)  + \frac{1}{2}f^\prime T_{{\rm matt}} }
{f\left(1 + \frac{3{f^\prime}^2}{2f}\right) } \,\,\,,
\end{equation}
where ${\cal L}_{\mbox{\boldmath{$n$}}} \Pi= 
\frac{1}{N}\partial_t \Pi + \frac{N^l}{N}\partial_l \Pi$ and 
$D_c Q^c = \,\!\partial_c Q^c + Q^c \partial_c ({\rm ln}\sqrt{h})$, 
here $h={\rm det}h_{ij}$ and $T_{{\rm matt}}= S_{{\rm matt}}-E_{{\rm matt}}$.
\bigskip

The Eqs. (\ref{Pi}), (\ref{evQ2}) and (\ref{evPi1}) are the evolution equations for the 
scalar-field variables. In fact, since one has introduced an evolution Eq. (\ref{evQ3}) for 
$Q_a$, formally one can also promote the definition of $Q_a$ Eq. (\ref{Q}) and the integrability condition 
$D_aQ_b=D_bQ_a$ as constraints. Taking the spatial components one has
\begin{eqnarray}
\label{constQ1}
Q_i-D_i\phi &=& 0 \,\,\,\,,\\
\label{constQ2}
D_{[i}Q_{j]} &=& 0 \,\,\,\,.
\end{eqnarray}

The system of evolution equations (both for the gravitational field and 
the scalar field) are to be completed with the evolution equations for the matter.
\bigskip

In summary, the Cauchy problem of STT can be sketched as follows: given 
the initial data set $(\Sigma_t, h_{ij}, K_{ij}, \phi,\Pi,Q_i)$ 
subject to the constraints Eqs. (\ref{CEHfSST}), (\ref{CEMfSST}), (\ref{constQ1}) and (\ref{constQ2}), 
the set of equations (\ref{K_ij}), (\ref{Pi}), (\ref{evQ2}), (\ref{EDEfSST}) and 
(\ref{evPi1}) provide the Cauchy development for the gravitational and the scalar 
field. Clearly this development can be unambiguously performed once the 
gauge (choice of coordinates) has been fixed. This is equivalent to fix the 
lapse and the shift. Note that these quantities appear in all the evolution equations. 
In the next section, the gauge choice for STT is proposed by generalizing several 
lapse and shift evolution equations that have been analyzed in the past.
\bigskip

Now, more comments about this new set of 3+1 equations for STT are in order.  
A nice feature of the constraint Eqs. (\ref{CEHfSST}) and (\ref{CEMfSST}) [see also (\ref{constQ1}) and (\ref{constQ2})] 
is that the lapse and the shift do not appear explicitly. These equations then 
constraint the initial data $(\Sigma_t, h_{ij}, K_{ij}, \phi, \Pi, Q_i) $ regardless 
of the coordinate choice. On the other hand, neither the constraints 
(\ref{CEHfSST}) and (\ref{CEMfSST}) [see also (\ref{constQ1}) and (\ref{constQ2})]  nor the dynamic Eq. (\ref{EDEfSST}) 
contain time derivatives of $\Pi$ or $Q_a$. 
Moreover, the system of 3+1 equations [notably  the dynamic Eqs. (\ref{EDEfSST}) and (\ref{evPi1}) ] are now of 
full first order (in time and space) in the scalar-field variables. 
This is to be contrasted with the pure gravitational sector of the 
equations where  SOSD of $h_{ij}$ appear when expanding the curvature $\,^3 R_{\,\,\,j}^i$
 or  in the explicit term $D^i D_j N$ for the lapse. More recently, new 
3+1 formulations (e.g. see \cite{KST,foBSSN} and references therein), provide a full first order system 
of the gravitational and gauge sectors which are better suited for numerical 
stability (see the discussion below).

In 
order to avoid any kind of pathologies 
one should only consider STT with $f>0$, particularly if one requires 
a mathematically well posed initial value problem (see Appendix A). For instance, the condition 
$f>0$ avoids the possibility for the appearance of potential singularities in 
Eqs. (\ref{CEHfSST}), (\ref{CEMfSST}) and (\ref{EDEfSST}). 
Notably in the terms with $1/f$ and $1/\left(1 +3{f^\prime}^2/(2f) \right)$. 
Those kind of singularities could preclude the existence of a Cauchy 
hypersurface 
\footnote{For instance, an initial data set corresponding to a static 
and spherically symmetric spacetime 
with $f= (1-4\pi G_0\phi^2/3)/(8\pi G_0)$ and $V(\phi)=0$, and without matter,
is the one associated with the well known solution of Bekenstein-Bronnikov-Melnikov-Bocharova 
\cite{BBMB}, where the spacetime has the geometry of the extreme Reissner-Nordstrom 
solution. In this solution, the scalar field is singular at the event horizon. 
Moreover, it has been argued that the Einstein 
field equations are not satisfied at the event horizon \cite{sudzan} 
(see also Refs. \cite{faraoni,abramo} and references therein 
for more considerations about this point and the conformal coupling). 
Those kind of pathologies could be avoided with the inclusion of matter 
(e.g. a cosmological constant $\Lambda$ with $V(\phi)= \lambda \phi^4$ and 
$\lambda= -2\pi \Lambda/9$ can give rise to regular black-hole solutions \cite{mtz}).}. 
Moreover, physically the condition $f>0$ ensures a positive definite 
effective gravitational constant $G_{\rm eff}$ [cf. Eq. (\ref{Geff})]. 

It is to note that if $f={\rm const.}$ the new set of 3+1 equations reduce to the 
standard 3+1 equations of GR with a minimally coupled scalar-field. On the other hand, as 
discussed in Appendix D, the full form of 3+1 equations of STT have passed successfully a self-consistency check that 
consists in computing the 3+1 conservation equations of the effective energy-momentum tensor 
of STT and showing that such equations together with the 3+1 field equations lead to the conservation of the 
energy-momentum tensor of matter alone.
\bigskip

The set of evolution equations (\ref{K_ij}) and (\ref{EDEf}) in the absence of 
matter fields and the scalar field (i.e, vacuum GR) cannot be straightforwardly 
submitted to an hyperbolicity analysis which is of main importance for establishing 
a well posed Cauchy problem. 
Only recently the 3+1 equations (including evolution equations for the gauge 
functions -see Sec. IV-) have been rewritten in full first order form, 
\begin{equation}
\label{PDE}
\partial_t {\vec u} + \mathbb{M}^i \nabla_i {\vec u} = {\vec S}({\vec u}) \,\,\,,
\end{equation}
where ${\vec u}$ represents collectively the fundamental variables (like the 
$h_{ij}$'s, $K_{ij}$'s, etc.), $\mathbb{M}^i$ is called the 
{\it characteristic} matrix of the system and ${\vec S}({\vec u})$ represents 
{\it source} terms which include only the fundamental variables 
(not their derivatives). This system of partial differential equations (PDE) is 
said to be {\it quasi-linear} since it is linear in the derivatives but in general 
it is non-linear in ${\vec u}$. A system of PDE of the form Eq. (\ref{PDE}) is 
said to be: 1) {\it weakly hyperbolic}, 2) {\it strongly hyperbolic}  
3) {\it symmetric hyperbolic} or 4) {\it symmetrizable hyperbolic}
 if the characteristic matrix $\mathbb{M}^i$ corresponding 
respectively to such cases satisfies: 1) Given any co-vector $s_i$, then $s_i \mathbb{M}^i$ 
has a real set of eigenvalues but not a complete set of eigenvectors, 
2) $s_i \mathbb{M}^i$ has a real set of eigenvalues and a complete set of eigenvectors (plus some bound
conditions), 3) $\mathbb{M}^i$ are symmetric, 4) $\mathbb{M}^i$ can be symmetrized.

Only strongly, symmetric and symmetrizable hyperbolic formulations (hereafter {\it hyperbolic 
formulations}) have a well posed Cauchy problem (see Ref. \cite{Reula} 
for a review). On the other hand, 
quasi-linear wave equations (QLWE) are hyperbolic and they have a 
well posed Cauchy problem \cite{Leray}. The so called 
{\it second order analysis} (as opposed to the first order one outlined above) 
consists in re-writing the evolution equations as QLWE. For instance, as it is 
well known, the vacuum Einstein's field equations 
reduce to a system of QLWE in the harmonic gauge.

After many years of using the 3+1-ADM equations as a ``work-horse'' for numerical 
studies, now it is known that in general Eqs. (\ref{K_ij}) and (\ref{EDEf}) (in vacuum)
even in its first order form like in Eq. (\ref{PDE}),
are not hyperbolic (only {\it weakly hyperbolic}) and that 
one needs to add some multiples of the constraints to turn the whole system 
hyperbolic \cite{KST} or at least to obtain a numerically stable system 
(see Ref. \cite{alcubierretal0}). 
Of course this also requires the use of suitable gauge choices (see Sec. IV). 
In this sense, it is expected that the new system of evolutions equations 
(\ref{K_ij}), (\ref{Pi}), (\ref{evQ2}), (\ref{EDEfSST}) and 
(\ref{evPi1}) [once put in first order form (\ref{PDE})] will not be a hyperbolic system either. Clearly the pure scalar-field sector is 
hyperbolic since Eq. (\ref{KG}) is a quasi-linear wave equation.
\bigskip

It is out of the scope of the present work to provide a full first order formulation
 of Eqs. 
(\ref{K_ij}), (\ref{Pi}), (\ref{evQ2}), (\ref{EDEfSST}) and 
(\ref{evPi1}) and its corresponding hyperbolicity analysis
(e.g., see Refs. \cite{KST,foBSSN} for the case of GR). 
Nonetheless, the new set of 3+1 equations 
is a first step towards this task. On the other hand, using the 
the covariant equations (\ref{Einst}) and (\ref{KG}), together with a generalization 
of the harmonic gauge for STT, it is shown in Appendix A 
that STT in the absence of matter have a well posed Cauchy problem.

\section{Slicing and Shift conditions}
As it was mentioned in previous section, the evolution of the 
gravitational and scalar fields can be performed without 
ambiguity once the lapse and the shift are fixed. This fixing amounts 
to impose coordinate conditions.

Let us remind that one of the most convenient coordinate (gauge) choices 
in GR which leads to a well posed initial value problem is the 
{\it De Donder} or {\it harmonic gauge} (HG) given by
\begin{equation}
\label{HG}
\Box x^a = 0= \frac{1}{\sqrt{-g}}\partial_c\left(\sqrt{-g} g^{ac}\right)= 
-g^{cd}\Gamma^{a}_{cd} \,\,\,\,.
\end{equation}
In the framework of the 3+1 formalism where 
the coordinates $x^a$ are chosen to coincide with $(t,x^i)$, that is, 
with the time global function that defines the spatial slices $\Sigma_t$ 
and the local spatial coordinates $x^i$ of the submanifold $\Sigma_t$, 
the HG leads to an evolution equation for the lapse 
$N$ and the shift $N^a$ (see below). 
For the analysis of the Cauchy problem of STT, a generalization of the 
harmonic gauge is proposed [hereafter {\it pseudo-harmonic gauge} (PHG)] in the following way
\begin{equation}
\label{STPHG}
\Box x^a = -\Theta \nabla^a {\cal F} \,\,\,\,,
\end{equation}
where ${\cal F}:= {\rm ln} F$ and the r.h.s of Eq. (\ref{STPHG}) is to 
be understood in the following way $\nabla^a {\cal F}= 
g^{ab}\nabla_b {\cal F}$. Here $\Theta$ is an arbitrary parameter. As analyzed in the 
Appendix A, for $\Theta=1$ this gauge leads in fact to a well posed initial value problem of STT. 
Moreover, in the linear-limit (see Appendix B) 
this gauge choice (with $\Theta=1$) generalizes the Lorentz-De Donder like gauge which leads
to wave equations for the spin-2 and spin-0 modes 
(gravitational and scalar waves propagating in ``Minkowski'' spacetime).

In numerical relativity, it is customary to chose the lapse and the 
shift conditions independently. This means that one does not necessarily 
impose the same condition Eq. (\ref{HG}) simultaneously for the time and spatial coordinates. 
It is in fact sometimes better to impose rather different equations 
for both the lapse (slicing condition) and the shift. One of the most popular family of slicings 
is the so-called Bona-Masso 
(BM) conditions \cite{BM} which can be written in four-dimensional 
notation as \cite{alcubierre1}:
\begin{equation}
\label{BM}
\Box t = \Upsilon_q n^b n^c \nabla_b \nabla_c t \,\,\,\,,
\end{equation}
where $\Upsilon_q= 1/q -1$ and $q=q(N)>0$ (i.e., $q$ is a positive but 
otherwise arbitrary function of the lapse). This BM condition provides the 
following evolution equation for the lapse (see Appendix C)
\begin{equation}
\label{EqN}
{\cal L}_{\mbox{\boldmath{$n$}}} N= 
\frac{1}{N}\partial_t N + \frac{N^l}{N}\partial_l N  = - N q(N) K\,\,\,\,.
\end{equation}
More recently, it was proposed a modification of the BM condition in the 
following way \cite{Yo,alcubierreetal1}:
\begin{equation}
\label{MBM}
\Box t = \frac{\Upsilon_q}{N} t^b n^c \nabla_b \nabla_c t 
-\nabla_c\left(\frac{N^c}{N^2}\right)\,\,\,\,.
\end{equation}

This modified Bona-Masso (MBM) condition is equivalent to
\begin{equation}
\label{EqNM}
\partial_t N= - N^2 q(N)\left( K + \frac{1}{N}D_c N^c\right) \,\,\,\,.
\end{equation}
This condition has the advantage that the lapse does not evolve if 
there exists a timelike Killing vector field $(\partial/\partial t)^a$. 
In that case $\partial_t h_{ij} \equiv 0$, 
[in particular for the determinant  
$\partial_t h= -2 h  \left( NK + D_c N^c\right)=0 $ -cf. Eq. (\ref{K_ij})- ] 
and therefore the r.h.s of Eq. (\ref{EqNM}) vanishes identically. 

For the BM or MBM, different choices of the function $q(N)$ have been analyzed 
both analytically and numerically \cite{BM,alcubierre1,alcubierreetal1}.

For the case of STT, the following lapse and shift conditions are considered 
(from which several gauges in GR can be recovered):
\begin{equation}
\label{STTBM}
 \Box x^a = A\Upsilon_{q_a} n^b  n^c \nabla_b \nabla_c x^a + B \delta^a_0\left[ 
\frac{\Upsilon_{q_a}}{N} t^b n^c 
\nabla_b \nabla_c x^a
-\nabla_c\left(\frac{N^c}{N^2}\right) 
- \Theta \frac{N^c}{N^2}Q_c {\cal F}^\prime\right]
- \Theta \nabla^a {\cal F} \,\,\,\,({\rm no\,\,summation\,\,over\,\,the\,\,index}\,\, a)\,\,.
\end{equation}
Here one allows to have four 
different functions $\Upsilon_{q_a}= 1/q_a -1$, with  $q_a=q_a(N)>0$, and 
where $A$ and $B$ are arbitrary coefficients. For instance, if $B=0$ and 
$A=1$, one has a STT generalization of the BM condition for $x^0=t$. 
If $A=0$ and $B=1$ one finds the STT generalization of the MBM condition 
for $x^0=t$. The pure GR gauge conditions are recovered trivially when 
${\cal F}={\rm const.}$

A straightforward but cumbersome algebraic manipulations using the 
3+1 formalism of Sec. III (see Appendix C), allows one to re-write Eq. (\ref{STTBM}) as 
the following evolution equation:
\begin{eqnarray}
\label{STTBMev}
 n^c\partial_c n^a - D^a({\rm ln}N) &=& \frac{1}{1+ \Upsilon_{q_a}(A+B\delta^a_t)}
\left\{ V^a + n^a K + B\delta^a_t 
\left[ \Upsilon_{q_a}\frac{N^c}{N}\left(\rule{0mm}{0.4cm}\partial_c n^a 
+ K^a_c\right) + 
\frac{1}{N^2}\left(\rule{0mm}{0.4cm}D_c N^c - N^c D_c({\rm ln}N)\right) \right] 
\right. \nonumber \\
&& \left. 
+ \Theta{\cal F}^\prime\left(Q^a-n^a\Pi + \delta^a_t \frac{B}{N^2}N^c Q_c\right) 
\right\} \,\,\,\,,
\end{eqnarray}
where one reminds $n^a=(1/N,N^i/N)$. Here 
$V^a:= \frac{1}{\sqrt{h}}\partial_c\left(\sqrt{h} h^{ac}\right)$, and it was used the 
following identity $\nabla^a{\cal F}=
{\cal F}^\prime\nabla^a\phi= 
{\cal F}^\prime\left(D^a\phi - n^a n^c \nabla_c\phi\right)
= {\cal F}^\prime\left(Q^a-n^a\Pi\right)$. Note for the time ``component'' $V^t\equiv 0$.

The time and the spatial parts of Eq. (\ref{STTBMev}) yield respectively the evolution equations for the 
lapse and shift:
\begin{equation}
\label{STTlapse}
\partial_t N + \frac{1+ A\Upsilon_{q_t} -B}{1 + \Upsilon_{q_t}(A+B)}N^j\partial_j N 
= - \frac{N^2}{1+ \Upsilon_{q_t}(A+B)}
\left\{ K + \frac{B}{N} D_c N^c - \Theta{\cal F}^\prime\left(\Pi- B\frac{N^c}{N}Q_c\right)
\right\} \,\,\,\,,
\end{equation}

\begin{equation}
\label{STTlapseshift}
n^c\partial_c\left(\frac{N^i}{N}\right)- D^i({\rm ln}N)
 = \frac{1}{1+ \Upsilon_{q_i}A}
\left\{ V^i + \frac{N^i}{N} K + 
\Theta{\cal F}^\prime\left(Q^i- \frac{N^i}{N}\Pi\right) \right\} \,\,\,\,.
\end{equation}
When the lapse and the shift equations (\ref{STTlapse}) and (\ref{STTlapseshift}) 
are imposed simultaneously and for the 
simple case where $B=0$ and the four functions $\Upsilon_{q_a}= \Upsilon_{q} $ 
are all equal then Eq. (\ref{STTlapseshift}) 
can be simplified using Eq. (\ref{STTlapse}) to yield
\begin{equation}
\label{STTshift}
 \partial_t N^i + 
N^j\partial_j N^i  - N D^i N = \frac{N^2}{1+ \Upsilon_{q}A}
\left\{\rule{0mm}{0.4cm} V^i + \Theta{\cal F}^\prime Q^i \right\} \,\,\,\,,
\end{equation}
which is valid if the lapse satisfies
\begin{equation}
\label{STTlapseB0}
\partial_t N + N^i\partial_i N 
= - \frac{N^2}{1+ \Upsilon_{q}A}
\left\{ K - \Theta{\cal F}^\prime \Pi \right\} \,\,\,\,.
\end{equation}

It is also usual to employ the variable ${\bar N}^i= N^i/N$ instead of the shift 
\cite{alcubierreetal2}. Then, 
Eq. (\ref{STTlapseshift}) reads
\begin{equation}
\label{STTsigma}
\partial_t {\bar N}^i + N{\bar N}^j\partial_j {\bar N}^i  - D^i N
  =  \frac{N}{1+ \Upsilon_{q_i} A}
\left\{\rule{0mm}{0.5cm} V^i + {\bar N}^i K +   
\Theta{\cal F}^\prime\left(\rule{0mm}{0.4cm}Q^i- {\bar N}^i\Pi\right) \right\} \,\,\,\,.
\end{equation}

Again, note from Eq. (\ref{STTlapseB0}) if $A=1$ and $\Upsilon_{q}= 1/q(N)-1$, 
one has a generalization of the BM slicing family for STT:
\begin{equation} 
\label{STTBMlapse}
\partial_t N + N^l\partial_l N  = - q(N)N^2 
\left(\rule{0mm}{0.4cm} K- \Theta{\cal F}^\prime \Pi\right) \,\,\,\,.
\end{equation}

In the same way, note from Eqs. (\ref{STTshift}) and (\ref{STTsigma}) that if $A=1$ 
and $\Upsilon_{q}= 1/q(N)-1$ and $\Upsilon_{q_i}= 1/p(N) -1$ one obtains 
the generalization for STT of the shift evolution equations proposed in Ref. 
\cite{alcubierreetal2}:
\begin{equation}
\label{STTBMshift}
\partial_t N^i + N^j\partial_j N^i  - N D^i N
= q(N) N^2 \left(\rule{0mm}{0.4cm} V^i + \Theta {\cal F}^\prime Q^i \right) \,\,\,\,,
\end{equation}

\begin{equation}
\label{STTBMsigma}
\partial_t {\bar N}^i + N{\bar N}^j\partial_j {\bar N}^i  - D^i N
 = p(N) N \left[\rule{0mm}{0.5cm} V^i + {\bar N}^i K 
+ \Theta {\cal F}^\prime\left(\rule{0mm}{0.4cm}Q^i- {\bar N}^i\Pi\right) \right] \,\,\,\,.
\end{equation}

On the other hand, note also from Eq. (\ref{STTlapse}) that for the case 
$A=0$ and $B=1$ and $\Upsilon_{q_t}= 1/q(N)-1$, 
one obtains a generalization of the MBM condition for STT:
\begin{equation} 
\partial_t N  = - q(N)N^2 \left[ K + \frac{1}{N}D_c N^c
- \Theta{\cal F}^\prime\left(\Pi -\frac{N^c}{N}Q_c\right)  \right] \,\,\,\,,
\end{equation}
which, in the presence of a timelike Killing vector field $(\partial/\partial t)^a$, 
also leads to a time-independent lapse (like in the original MBM 
condition) because of the arguments given before and also because 
in that case $\partial_t \phi \equiv 0$ which
by virtue of Eq. (\ref{Pi}) then leads to $N\Pi - N^c Q_c= 0$.

Finally, if $q(N)=1$, then from Eqs. (\ref{STTBMlapse}) and (\ref{STTBMshift}) one 
recovers the PHG Eq. (\ref{STPHG}) here in terms of the lapse and shift

\begin{eqnarray} 
\label{STTHlapse}
&& \partial_t N + N^l\partial_l N  = - N^2 
\left(\rule{0mm}{0.4cm} K- \Theta{\cal F}^\prime \Pi\right) \,\,\,\,,\\
\label{STTHshift}
&& \partial_t N^i + N^j\partial_j N^i  - N D^i N
= N^2 \left(\rule{0mm}{0.4cm} V^i + \Theta{\cal F}^\prime Q^i \right) \,\,\,\,.
\end{eqnarray}

In Appendix B the linear limit of Eqs. (\ref{STTHlapse}) and 
(\ref{STTHshift}) will be taken and show that this pair of equations generalizes 
the Lorentz-De Donder gauge of GR to the case of STT. In pure GR (${\cal F}= {\rm const.}$), 
the set of evolution Eqs. (\ref{STTHlapse}) and (\ref{STTHshift}) reduce to 
the harmonic gauge that was considered in the past by Smarr \& York \cite{S&Y78} 
in terms of the lapse and shift \footnote{One should notice a sign error in 
the second of Eqs. (6.4) for the shift in the Smarr \& York article \cite{S&Y78} 
(it should read ``$-$'' instead of ``$+$''
 in front of the term $\gamma^{ij}\Gamma^{i}_{jk}$ ). For the purpose of comparison with that article or others which use a different 
sign convention and notation for the shift, one should take into account 
the following $N^i=-\beta^i$.}. 

In the case of GR, the BM and MBM conditions (given a prescribed shift) 
have been shown to be consistent 
with hyperbolic formulations \cite{BM,alcubierreetal1}. Alternatives to the 
slicing conditions are the use of a {\it densitized lapse} 
$\Xi={\rm ln}[Nh^{-\beta}]$ \cite{KST} 
(where $\Xi(t,x^i)$ is an arbitrary but {\it a priori} known 
function of the spacetime coordinates and $\beta$ is 
a constant; a positive $\beta$ is favored for strong hyperbolicity) 
or to take the lapse $N(h,t,x^i)$ as a prescribed function of the determinant $h$ 
of the 3-metric and the spacetime coordinates \cite{sarbach} ($\partial_h N >0$ is 
required for strong hyperbolicity).

On the other hand, 
it is important to emphasize that in pure GR (not even to mention the STT) 
only a few shift evolution equations have been 
tested numerically in the past or included in the analysis of hyperbolicity 
\cite{Lindblom,Beyer,BonaPal}.
A particularly simple shift is a null one 
(i.e., $N^i=0$, $\partial_t N^i=0$; gauge not included 
as a special case of the above shift conditions unless one considers a 
constrained initial lapse). However, this shift 
choice is not always free of numerical pathologies. Shift conditions have not 
been explored thoroughly, and only empirical choices have proved to be 
successful in numerical experiments. 
The shift equation (\ref{STTBMsigma}) in GR (${\cal F}= {\rm const.}$) have been analyzed 
numerically for very special cases \cite{alcubierreetal2}.

There also exists gauges leading to elliptic equations rather than wave equations for 
$N$ and $N^i$. One is the {\it maximal slicing condition} ($K=0$, $\partial_t K=0$) 
which in GR would correspond to take $q(N)$ arbitrarily large [cf. Eq. (\ref{EqN})]. 
Then Eq. (\ref{EDK2}) provides an elliptic equation for $N$. An elliptic equation 
for $N^i$ was considered by Smarr \& York \cite{S&Y78} (which includes the so-called 
{\it minimal distortion gauge}). Both elliptic equations were shown to be the 
analogous in GR of the {\it radiation gauge} in electrodynamics. It would be 
interesting to generalize those conditions in STT. For, in Eq. (\ref{STTBMsigma}) 
taking $q(N)$ arbitrarily large would lead to $K=\Theta {\cal F}^\prime \Pi$ which for 
$\Theta \neq 0$ and 
in view of the evolution Eqs. (\ref{EDKSTT}) and (\ref{evPi1}) one would obtain a 
complicated elliptic equation for $N$. 

In any case, 
a full and new numerical exploration of the above gauges (or others) is to be performed in order 
to test their usefulness in the context of STT.

\section{Specific applications}
The new system of 3+1 equations presented in Sec. III when applied 
to specific spacetimes coincides with the system of equations analyzed in the past 
\cite{us,salgado2,salgado3,salsudnuc} and which were derived from the covariant field 
equations in a rather circuitous way in order to recast them into first order form. 
Below, some details about some of these specific situations are discussed.
\bigskip

{\bf Cosmology}. In the case of an isotropic and homogeneous spacetime (Friedmann-Robertson-Walker 
cosmology) with a combination of two non-interacting perfect fluids (representing photons and 
cold matter), the 3+1 equations of Sec. III reduce to those of Refs. \cite{us,salgado3} for the class of 
STT where $F(\phi)= 1+ 16\pi G_0 \xi \phi^2$ [$\xi>0$; recall $f(\phi)= F/(8\pi)$] and $V(\phi)= m^2\phi^2$. In particular, 
the Hamiltonian constraint Eq. (\ref{CEHfSST})
leads to the Friedmann equation relating the Hubble parameter with the 
total energy-density of the Universe. For this case, the extrinsic curvature is related to the 
Hubble expansion as $K= -3\dot a/a$ [$a(t)$ being the scale-factor]
and $^3 R= 6k/a^2$ (where $k=\pm 1,0$). 
On the other hand, the dynamic equations (\ref{EDEfSST}) lead to a single equation for the acceleration of the 
Universe ($\ddot a$). As shown in Refs. \cite{us,salgado2}, 
the two first order equations [one for $\dot a$ resulting from Eq. (\ref{K_ij})] 
and the other for $\dot K$ [resulting from Eq. (\ref{EDKSTT})] together with the 
Eqs. (\ref{Pi}) and (\ref{evPi1}) for the scalar-field can be integrated numerically. 
The momentum constraint Eq. (\ref{CEMfSST}), 
as well as the constraints Eqs. (\ref{constQ1}) and (\ref{constQ2}), are satisfied 
trivially in this case. 
The Hamiltonian constraint can be used 
at every time step to monitor the numerical accuracy of the results \footnote{Here it is important to 
mention a typographical mistake committed in Eqs. (34) and (A19) of Ref.
M. Salgado, D. Sudarsky, and H. Quevedo, \Journal{\PRD}{53}{6771}{1996} that concerns the sign of the 
Universe's spatial curvature given in terms of the total-energy density. However, the correct sign was taken 
into account in the numerical code and hence the results are not affected.}. Regarding the matter, the matter 
conservation equations together with the equation of state (EOS) for photons and dust lead to the standard 
quadrature for the total energy-density of matter $\rho_{\rm matt}= c_1/a^4 + c_2/a^3$ 
(where $c_1$, $c_2$ are 
constants indicating the content of photons and ``dust'' -baryons and dark matter- 
at a given cosmic time). 
This set of equations were successful for analyzing 
the evolution of the Universe by taking as ``initial'' conditions the conditions 
of the Universe today and integrating the equations backwards in time. The initial conditions for the 
scalar-field and the value of the NMC constant $\xi$ can be chosen so as to satisfy several local constraints \cite{us}. The 
concrete STT that results is completely rigid (falsifyable) in the sense that  all the parameters are fixed for 
once and for all and that the specific STT should account for all the observations (which in fact 
give raise to a larger number of algebraic constraints than the available parameters of the theory).  
Among these observations 
one can mention the primordial nucleosynthesis (specially that of ${\rm He}^4$ 
which restricts severely the 
expansion rate of the Universe at early times), the type-I supernova distance-relation, the 
Cosmic Microwave Background measurements, etc.

At this point it is worth mentioning that a remarkable number of investigations in cosmology using the Brans-Dicke theory or 
other STT exist in the literature. It is completely out of the scope of this work to comment about 
each of them. Here the aim is simply to make contact between the equations of Sec. III 
and their use in cosmology.
\bigskip

{\bf Static and spherically symmetric spacetimes}. In the case of a spacetime admitting a static time-like Killing 
vector field and $SO(3)$ symmetry, and also for the class of STT with $F(\phi)= 1+ 16\pi G_0 \xi \phi^2$ ($\xi>0$), the 3+1 equations of Sec. III reduce to those of Refs. \cite{salgado2,salsudnuc}. In those references, for the matter 
it was taken a perfect-fluid representing a neutron star. In this case, one is basically dealing with 
solving the constraints (the Hamiltonian constraint), the equation for the lapse (the norm of the 
time-like Killing vector field) and the Eqs. (\ref{Q}) and  (\ref{evPi1}) for 
the scalar-field. In this case the quantity $D_i Q^i$ which appears in the Hamiltonian 
constraint Eq. (\ref{CEHfSST}) can be 
replaced in terms of simpler quantities using the static limit of Eq. (\ref{evPi1}). 
The boundary conditions imposed in Refs. \cite{salgado2,salsudnuc}
were such as to obtain an asymptotically flat spacetime and fields regular at the origin. This analysis 
led to the recovery of the phenomenon of spontaneous scalarization discovered by Damour \& Esposito-Far\`ese \cite{damour3}, but now using realistic EOS and in the Jordan frame. Moreover, 
a further 
analysis along this line can be found in Ref. \cite{salsudnuc} 
(this deals mainly with the energy-conditions).
\bigskip

Within the context of static and spherically symmetric spacetimes, 
it is perhaps worth mentioning some of the situation of black holes in STT. In the early 
seventies it was somehow a surprise to find that for the conformal coupling
$F(\phi)= 1-4\pi G_0\phi^2/3$ and $V(\phi)=0$ (and without matter) the field 
equations admitted an exact asymptotically flat static and spherically symmetric 
(AFSSS) solution (the Bekenstein-Bronnikov-Melnikov-Bocharova solution) \cite{BBMB}
which corresponds actually to the extreme Reissner-Nordstrom geometry, except 
that the scalar-field replaces the electric field. Despite the pathologies 
of the scalar field at the horizon (it is singular there), it was considered 
as a genuine solution, until very recently. Sudarsky \& Zannias \cite{sudzan} 
carefully reanalyzed the issue and showed that in fact due to this pathology 
the ``solution'' does not hold at the horizon. The lack of regularity 
of the scalar field at the horizon (a standard requirement for the existence and uniqueness 
of solutions) was then the responsible of this anomaly (see also Ref. \cite{salgado4}). 
Actually, the regularity at the horizon is also a crucial condition for establishing the 
opposite situation: the non-existence of scalar-hair (the so-called no hair theorems) 
\cite{nohair}. These theorems apply for STT with a minimally coupled 
scalar-field and non-negative potentials $V(\phi)\geq 0$ for AFSSS. For the case 
of NMC, no-hair theorems also exist for certain class of STT \cite{nomin}.

When the condition $V(\phi)\geq 0$ is abandoned, hairy back holes have shown (at least 
numerically) to exist \cite{nucsal}. Moreover, by introducing a cosmological 
constant, it is possible to find scalar-hairy black holes in asymptotically de-Sitter 
and anti-de Sitter 
spacetimes with scalar-fields coupled both minimally and no-minimally \cite{mtz,bhads}.

Nevertheless, for asymptotically flat spacetimes and with $V(\phi)\geq 0$ or $V(\phi)\equiv 0$, 
there is still many open questions about the existence of regular black hole solutions in STT.

In many aspects the new 3+1 equations of Sec. III can be useful for the analysis of 
stationary and static spacetimes. 
In those cases, many degrees of freedom cancel out and the equations 
simplify considerably. For instance, in the case of static spacetimes, 
$K_{ij}\equiv 0 \equiv \Pi$. The momentum 
constraint Eq. (\ref{CEMfSST}) reduces to $0\equiv 0$.
In static and spherical symmetric spacetimes 
the simplification is even greater since one can chose $N^i\equiv 0$. In fact, one has 
only to solve an elliptic equation for the lapse [cf. Eq. (\ref{EDKSTT}); or 
alternatively a first-order equation for the lapse which results from the term $D^iD_j N$ 
when using the angular components of the dynamic Eq. (\ref{EDEfSST})],  
the two first-order equations for the scalar field [cf. Eqs. 
(\ref{Q}) and (\ref{evPi1})], and the 
Hamiltonian constraint for one metric potential (e.g., $g_{rr}$). Moreover, 
when employed the parametrization $g_{rr}= (1-2m(r)/r)^{-1}$, 
the Hamiltonian constraint provides a very simple ordinary differential equation for 
$m(r)$. Indeed and as commented above, when taking into account a perfect fluid, it is 
such a set of equations that were used in Refs. \cite{salgado2,salsudnuc} to treat 
the phenomenon of spontaneous scalarization.
\bigskip

{\bf Dynamical spacetimes}. The dynamics of spacetimes in STT have been analyzed in the 
past only under very symmetric cases. In Brans-Dicke theory the spherically symmetric 
collapse of dust to a black-hole formation (Oppenheimer-Snyder collapse) 
has been analyzed by 
Sheel, Shapiro \& Teukolsky \cite{SST1,SST2}. These are as far as the author is aware, 
one of the few numerical studies where a STT is analyzed in the Jordan frame. 
A 3+1 decomposition of the equations are considered (but 
only for the special case of spherical symmetry). The slicing condition employed there 
was {\it maximal slicing}, and the spatial coordinates are of isotropic type 
(with a no-null shift).

A similar study performed a little bit earlier was carried out by Shibata, Nakao and 
Nakamura \cite{SNN} (see also \cite{HCNN}), also in BD theory and for the Oppenheimer-Snyder collapse.
This study, however, used the Einstein frame for which the standard 3+1 formalism was 
applied. The maximal slicing condition was adopted together with a null shift. 
The scalar-wave signal resulting from the dynamics was confronted with the sensibilities 
of interferometric detectors like LIGO.

More recently, Novak \cite{novak} performed several studies in 
spherically symmetric neutron-star models 
within certain class of STT. In Novak's analysis the Einstein frame 
was used and the corresponding 3+1 decomposition (which is the standard one) was implemented for this particular 
case. The coordinates used there correspond to the simple radial-polar-slicing gauge. 
The physical situation was to describe the dynamical transition to spontaneous-scalarization 
and the further gravitational collapse to a black-hole formation. In Novak's 
works, the scalar-wave emission 
was analyzed and confronted with the sensibility of current detectors. 

Using a perturbative analysis, STT have been also analyzed in neutron-star oscillations 
and the emission and detection of scalar-waves have been confronted with the imprint left on 
the gravitational wave (spin-2) spectrum \cite{kokkotas}.

Except for the above studies (in spherical symmetry or using perturbation theory) 
up to the author's knowledge, there is no further 
attempt in STT for studying the dynamics of spacetimes with less symmetries
 or with another kind of matter (different from a perfect-fluid).

Actually, one of the scopes 
for a future work is to analyze the dynamical transition to spontaneous scalarization 
and gravitational collapse with scalar-wave emission but for the case of boson stars 
(as opposed to neutron stars). In this case 
the matter is represented by a complex scalar-field $\psi$ with the 
following energy-momentum tensor:
\begin{eqnarray}
T_{ab}^{\rm matt} &=& (\nabla_a \psi^*)(\nabla_b \psi)+ 
(\nabla_b \psi^*)(\nabla_a \psi)  - g_{ab}
\left[ \frac{1}{2}|\nabla \psi|^2 + V_\psi(|\psi|^2) \right ]   \,\,\,,\\
\label{bospot}
V_\psi(\psi^*\psi)&=& m_\psi \psi^* \psi + \lambda(\psi^* \psi)^2 \,\,\,\,,
\end{eqnarray}
with the corresponding 3+1 variables given as
\begin{eqnarray}
\label{Ematp}
E_{\rm matt} &=& \frac{1}{2}\left( |\Pi_\psi|^2 + |Q_\psi|^2 \right) + V_\psi(|\psi|^2) \,\,\,,\\
S_{\rm matt}^{ab}&=& Q^{*a}_\psi Q^b_\psi + Q^{*b}_\psi Q^a_\psi -  
h^{ab}\left[ \frac{1}{2}\left( |Q_\psi|^2- 
|\Pi_\psi|^2 \right) + V_\psi(|\psi|^2) \right] 
\label{Smatp}\,\,\,,\\
\label{TrSmatp}
S^{\rm matt} &=& \frac{1}{2}\left(3|\Pi_\psi|^2 - |Q_\psi|^2 \right) -3 V_\psi(|\psi|^2) \,\,\,,\\
\label{S-Ematp}
S_{\rm matt} - E_{\rm matt} &=& |\Pi_\psi|^2 - |Q_\psi|^2 - 4 V_\psi(|\psi|^2)
\,\,\,,\\
\label{Jibos}
J_a^{\rm matt} &=& \Pi_\psi Q_a^{*\psi} + \Pi^*_\psi Q_a^\psi \,\,\,.
\end{eqnarray}
where
\begin{eqnarray}
\label{Qpsi}
Q_a^\psi &:=& D_a\psi \,\,\,,\\
\Pi^\psi &:=& {\cal L}_{\mbox{\boldmath{$n$}}} \psi= n^a\nabla_a\psi 
=\frac{1}{N}\left(\partial_t \psi + N^a Q_a^\psi \right) \,\,\,,
\label{Pipsi}
\end{eqnarray}
and $Q_a^{*\psi}$ and $\Pi^{*\psi}$ are obtained by conjugation from the above variables. Here $|\,\,|$ indicates 
the norm of the complex fields.

From the point of view of the numerics, this is much more simpler than the realistic 
and difficult analysis of neutron stars. In this case, the problem involves an extra 
Klein-Gordon equation 
\begin{equation}
{\Box \psi} = \partial_{\psi^*} V_\psi(\psi^*\psi) \,\,\,,
\end{equation}
[which can be easily written in 3+1 form like Eq. (\ref{evPi1})] instead of the 
relativistic Euler equation for a perfect fluid. This problem involves therefore 
two scalar-fields ($\phi$ and $\psi$).

In the Einstein frame and for the stationary situation, the phenomenon of spontaneous scalarization was shown to exist also 
in boson stars \cite{whinnett}. Nevertheless the dynamical analysis to that 
``phase transition'' has not been yet performed.   
\bigskip

\section{Outlook and Conclusions}
In this paper the field equations of scalar-tensor theories of gravity in the Jordan-frame representation have been 
recasted in a 3+1 form. This new system of equations shows that   
the Cauchy (initial value) problem is well formulated in that frame.
This result opens the doors to several possibilities, both analytical and numerical. 
For instance, it would be interesting to write the 3+1 equations in full first-order 
form in order to perform a hyperbolic analysis along the lines of recent research 
in pure GR. In this direction, several gauge conditions for STT 
which generalize gauge conditions used in GR, have been proposed. These latter 
correspond to evolution equations for the lapse and the shift vector.

Regarding the applications, a future plan is to use the formalism presented here 
to the case of several astrophysical phenomena. One of this is the numerical 
analysis of the transition to spontaneous scalarization and the collapse of 
boson stars. This would be similar but in many respects simpler than the 
analysis performed in neutron stars \cite{novak}.

On the other hand, when applied to specific spacetimes (for cosmology and 
spherically symmetric spacetimes) accompanied with perfect-fluid models, 
the new system of equations was shown to reduce to 
the equations analyzed numerically in the past 
by the author and colleagues \cite{us,salgado2,salsudnuc}. This illustrates that the 3+1 formulation derived 
here is successful in producing concrete results suitable for confrontation against 
observations.

In this paper, it was also shown from the covariant form of the equations, that 
STT have a well posed Cauchy problem using a generalization of the harmonic 
coordinates (Appendix A). For completeness, the linear and the Newtonian 
limits of STT were re-analyzed in Appendix B.

\begin{acknowledgments}
I'm indebted to M.Alcubierre for his comments and suggestions that improved the 
final form of the 3+1 equations. I wish to thank D.Sudarsky for proposing several 
lines of research based on this work.
 
This work has been supported by DGAPA-UNAM, grant Nos. IN122002 and IN119005.
\end{acknowledgments}

\appendix

\section{On the well-posedness of the Cauchy problem of STT}
In the following, the proof of the well-posedness of the Cauchy problem 
of the system of Eqs. (\ref{Einst}) and (\ref{KG}) in the absence of matter will be considered. 
To do so, it will be convenient to chose a generalization of the {\it harmonic coordinates} 
$x^a$ given by the condition
\begin{equation}
\label{Hgauge1}
{\tilde H}^a:= \Box x^a - {\cal G} \nabla^a\phi=0 \,\,\,,
\end{equation}
where ${\cal G}$ is a function of the scalar-field which is to be determined so that 
the Einstein equations acquire a desired form. 
Here $\nabla^a\phi=g^{ab}\nabla_b {\cal \phi}$ and as in Sec.IV, is to 
be understood in the sense of four functions in this particular coordinate system. As before,
\begin{equation}
\label{HG2}
\Box x^a = \frac{1}{\sqrt{-g}}\partial_b\left[\sqrt{-g} g^{ab}\right] 
= \partial_b g^{ab} + \frac{1}{2} g^{ab} g^{cd}\partial_{b}g_{cd}= -g^{cd}\Gamma^{a}_{cd}\,\,\,.
\end{equation}
Now, Eq. (\ref{Einst}) can be written in terms of $T:= T^{a}_{\,\,a}$, as 
\begin{equation}
\label{Einstalt}
R_{ab}= 8\pi G_0 \left( T_{ab}- \frac{1}{2}g_{ab} T\right)\,\,\,.
\end{equation}
Using the explicit expression for the Ricci tensor in terms of the metric and 
Eqs. (\ref{Hgauge1}) and (\ref{HG2}), one has
\begin{eqnarray}
\label{Ricci}
R_{ab}&=& R_{ab}^{\tilde H} - {\cal G} \partial^2_{ab} \phi 
- g_{c(a}\partial_{b)} {\tilde H}^c  = 
8\pi G_0 \left( T_{ab} - \frac{1}{2}g_{ab} T\right)\,\,\,\,, \\
\label{RicciH}
R_{ab}^{\tilde H}&:=& -\frac{1}{2} g^{cd}\partial^2_{cd}g_{ab} 
+  \mathscr{A}_{ab} (g,\partial g,\phi,\partial\phi)\,\,\,\,,
\end{eqnarray}
where $\mathscr{A}_{ab} (g,\partial g,\phi, \partial\phi)$ are 
non linear functions of the metric, of the scalar field, 
and their first derivatives (indices are omitted for brevity in the arguments). 

On the other hand the r.h.s of Eq. (\ref{Einstalt}) reads
\begin{equation}
\label{Riccirhs}
R_{ab} = \frac{f^\prime}{f}\partial^2_{ab}\phi + 
\mathscr{B}_{ab} (g,\partial g,\phi, \partial\phi) \,\,\,.
\end{equation}
where Eq. (\ref{KG}) was used to replace $\Box \phi$ in terms 
of its r.h.s which contains no higher than first order derivatives of the scalar field, 
and also the relation $T= S-E$ was employed (without the 
matter terms) which is given in general by Eq. (\ref{S-Esst}). 

Therefore Eq. (\ref{Einstalt}) reduces to 
\begin{equation}
\label{Einstred0}
R_{ab}^{\tilde H} - g_{c(a}\partial_{b)}{\tilde H}^c
 = \left({\cal G}+\frac{f^\prime}{f}\right)\partial^2_{ab}\phi +   
 \mathscr{B}_{ab} (g,\partial g,\phi, \partial\phi)  \,\,\,.
\end{equation}
The convenient choice 
\begin{equation}
{\cal G}= -\frac{f^\prime}{f}\,\,\,,
\end{equation}
leads then to 
\begin{equation}
\label{Einstred}
R_{ab}^{\tilde H} - g_{c(a}\partial_{b)}{\tilde H}^c
 =\mathscr{B}_{ab} (g,\partial g,\phi, \partial\phi)  \,\,\,.
\end{equation}
The gauge Eq. (\ref{Hgauge1}) reads then \footnote{One recognizes in Eq. (\ref{Hgauge2}) the 
gauge $ \Box_* x^a=0$ where $\Box_*$ is the D'Alambertian with respect to the conformal metric 
$g_{ab}^* = F(\phi) g_{ab}$.}
\begin{equation}
\label{Hgauge2}
{\tilde H}^a= \Box x^a + \frac{f^\prime}{f} \nabla^a\phi= 
 \Box x^a + \nabla^a {\cal F}= 0 \,\,\,,
\end{equation}
where ${\cal F}= {\rm ln} F$, which is exactly the same as the PHG (\ref{STPHG}) for $\Theta=1$.

Now, assuming that ${\tilde H}^c=0$ can be 
maintained during the evolution (as shown below, this is indeed the case), then 
\begin{equation}
\label{Einstred2}
R_{ab}^{\tilde H} 
 = \mathscr{B}_{ab} (g,\partial g,\phi, \partial\phi)  \,\,\,.
\end{equation}
From Eqs. (\ref{Einstred2}) and (\ref{RicciH}) one has
\begin{equation}
\label{Einstred3}
g^{cd}\partial^2_{cd}g_{ab}= \mathscr{C}_{ab} (g,\partial g,\phi,\partial\phi)\,\,\,.
\end{equation}
On the other hand Eq. (\ref{KG}) together with the gauge Eq. (\ref{Hgauge2}), yields
\begin{equation}
\label{KGred}
g^{cd}\partial_{cd}^2 \phi = \mathscr{D}(g,\phi, \partial\phi)  \,\,\,\,\,.
\end{equation}

The non-linear functions $\mathscr{A}_{ab}$, $\mathscr{B}_{ab}$, $\mathscr{C}_{ab}$, 
and $\mathscr{D}$ contains the explicit dependence of $\phi$ from $f,f^\prime,f^{\prime\prime}$ 
and $V$, $V^\prime$, and there appears also terms like $1/f$. 
Here it is assumed that the function $f$ is a positive definite smooth function, 
and that $V(\phi)$ is also smooth and therefore that 
$\mathscr{C}_{ab}$ and $\mathscr{D}$ are smooth functions of its variables. 
According to a theorem by Leray \cite{Leray} a system of 
quasilinear partial differential equations like 
(\ref{Einstred3}) and (\ref{KGred}) has a well posed initial 
Cauchy value problem in the sense that 
given some initial data in $\Sigma_t$ and assuming $(M,g_{ab})$ to be globally hyperbolic, 
the system has a unique solution and the solution depends 
continuously on the initial data (see Theorem 10.1.3 of Wald \cite{wald}).
\bigskip

What remains to be proved is that ${\tilde H^a}=0$ in a neighborhood of $\Sigma_t$ 
given ${\tilde H^a}=0$ and $\partial_t{\tilde H^a}=0$ on $\Sigma_t$. 
First,  ${\tilde H^a}=0$ on $\Sigma_t$ simply provides 
the initial values $\partial_t N$ and $\partial_t N^i$ on $\Sigma_t$ given 
the set $(h_{ij},K_{ij},\phi,\Pi)_{\Sigma_t}$ that satisfy the constraints 
$n^a R_{ab}= 8\pi G_0 n^a\left(T_{ab}-\frac{1}{2} g_{ab} T\right)$. Moreover, if the initial values are chosen as to satisfy 
the ``reduced'' constraints 
$n^a R_{ab}^{\tilde H}= n^a \mathscr{B}_{ab} (g,\partial g,\phi, \partial\phi)$ 
on $\Sigma_t$, this implies 
$\left(n^a g_{c(a}\partial_{b)}{\tilde H}^c\right)_{\Sigma_t}=0$, 
which in turn leads to $(\partial_t {\tilde H}^a)_{\Sigma_t} =0$. In this way 
the pseudo-harmonic gauge (\ref{Hgauge2}) and its time derivative are achieved initially. 

Finally, one can consider the effective Einstein tensor, ${\bar G}_{ab}:= G_{ab} - T^{STT}_{ab}$, where
$ G_{ab}$ is the usual Einstein tensor and $T^{STT}_{ab}$ is the energy-momentum tensor (\ref{effTmunu}) in the absence of 
matter. The gravitational field equations can be written as
\begin{equation}
\label{EinsSTT}
{\bar G}_{ab}=8\pi G_{\rm eff} T^{\rm matt}_{ab}\,\,\,\,,
\end{equation}
which in the absence of matter and in the PHG are equivalent to Eq. (\ref{Einstred2}). Now, the key point in the proof for the PHG to be maintained during the 
evolution will be the use of the effective Bianchi identities 
$\nabla^b ({\bar G}_{ab}/G_{\rm eff})=0$. These latter result from the usual Bianchi identities and 
from the field equations. In fact, as mentioned in the introduction, these identities in the presence of matter lead to the conservation 
of the energy-momentum tensor of matter alone. However, in the absence of matter, which is the case at hand,
the identities reduce simply to $\nabla^b {\bar G}_{ab}=0$. One has then that the effective Einstein tensor can be written in arbitrary coordinates 
in terms of ${\bar R}_{ab}^{\tilde H}$ [which is given by Eq. (\ref{RicciH}) 
with $\frac{1}{2}\mathscr{C}_{ab} (g,\partial g,\phi,\partial\phi)$ instead of  
$\mathscr{A}_{ab} (g,\partial g,\phi,\partial\phi)$; 
cf. Eq. (\ref{Einstred3})] and ${\tilde H}^a$ given by Eq. (\ref{Hgauge2}) as follows,
 \begin{equation}
{\bar G}_{ab}= {\bar R}_{ab}^{\tilde H} -\frac{1}{2}{\bar R}^{\tilde H} g_{ab} -
g_{c(a}\partial_{b)}{\tilde H}^c +\frac{1}{2}g_{ab}\partial_c {\tilde H}^c \,\,\,\,.
\end{equation}
Now, the use of $\nabla^b {\bar G}_{ab}=0$ and when Eqs. (\ref{Einstred3}) and (\ref{KGred}) are satisfied, leads to 
\begin{equation}
\nabla^b {\bar G}_{ab}= -\frac{1}{2} g_{ab} g^{cd} \partial_{cd}^2 {\tilde H}^b
+ \mathscr{E}_a(\partial{\tilde H},...)= 0 \,\,\,,
\end{equation}
where $\mathscr{E}_a(\partial{\tilde H},...)$ 
includes lower order terms linear in $\partial {\tilde H}$. From the above 
equation one easily writes
\begin{equation}
g^{cd} \partial_{cd}^2 {\tilde H}^a
= \mathscr{F}^a(\partial{\tilde H},...) \,\,\,,
\end{equation}
for which there exists a well-posed initial Cauchy value problem. Therefore, this equation ensures that
${\tilde H}^a\equiv 0$ in the region where solutions of Eqs. (\ref{Einstred3}) 
and (\ref{KGred}) exist,  if $\left({\tilde H}^a\right)
_{\Sigma_t}=0=\left(\partial_t {\tilde H}^a\right)_{\Sigma_t}$. 
In this way the pseudo-harmonic gauge is completely achieved.

\section{The linear limit of STT}

\subsection{The linear limit ``covariant'' approach}

The linear limit of STT has been analyzed in the past by many authors 
(see Refs. \cite{scalwaves}). For completeness, this limit is reanalyzed here and 
the point of departure will be the field equations in the Jordan frame. 
As usual, one considers first order perturbations of the Minkowski spacetime: 
\begin{eqnarray}
g_{a b } &\approx &\eta _{a b }+\epsilon \gamma _{a b }\,\,\,\,,
\\
T_{a b } &\approx &T_{a b }^{0}+\epsilon \tilde{T}_{a b
}\,\,\,\,, \\
\label{scalaproxi}
\phi &\approx &\phi _{0}+\epsilon \tilde{\phi}\,\,\,\,, \\
F(\phi ) &\approx &F_{0}+\epsilon F_{0}^{\prime }\tilde{\phi}\,\,\,\,, \\
\partial _{\phi }F(\phi ) &\approx &F_{0}^{\prime }+\epsilon F_{0}^{\prime
\prime }\tilde{\phi}\,\,\,\,, \\
\partial _{\phi \phi }^{2}F(\phi ) &\approx &F_{0}^{\prime \prime }+\epsilon
F_{0}^{\prime \prime \prime }\tilde{\phi}\,\,\,\,, \\
V(\phi ) &\approx &V_{0}+\epsilon V_{0}^{\prime }\tilde{\phi}\,\,\,\,, \\
\label{scalaproxf}
\partial _{\phi }V(\phi ) &\approx &V_{0}^{\prime }+\epsilon V_{0}^{\prime
\prime }\tilde{\phi}\,\,\,\,.
\end{eqnarray}
where $\epsilon \ll 1$ and the knot indicates quantities at zero order. In
the 4+0 covariant formulation one can introduce the combination 
\begin{eqnarray}
\tilde{\gamma}_{a b }:= &&\bar{\gamma}_{a b }+\kappa \eta _{a b }
\tilde{\phi}\,\,\,\,,  \label{tilgamma} \\
\label{bargamma}
\bar{\gamma}_{a b }:= &&\gamma _{a b }-\frac{1}{2}\eta _{a b
}\gamma \,\,\,\,,
\end{eqnarray}
where $\gamma =\gamma _{\,\,\,a }^{a }$ and $\kappa $ is a gauge
constant to be fixed later in order to simplify the equations. It is to be 
reminded that four dimensional indices are lowered or raised with the flat metric 
$\eta_{ab}$ (or $\eta^{ab}$).

The resulting linearized Einstein Eq. (\ref{Einst}) is
\begin{eqnarray}
\tilde{G}_{a b }&=& \partial ^{c }\partial _{(b }\tilde{\gamma}_{a
)c }-\frac{1}{2}\Box _{\eta }\tilde{\gamma}_{a b }-\frac{1}{2}\eta
_{a b }\partial ^{c }\partial ^{d}\tilde{\gamma}_{cd}\nonumber \\
&=& 8\pi G_{0}\tilde{T}_{a b }+\kappa \left( \partial _{a b
}^{2}\tilde{\phi}-\eta _{a b }\Box _{\eta }\tilde{\phi}\right) \,\,\,\,.
\label{einstlin}
\end{eqnarray}
where $T_{a b }^{0}=0$ results from the self-consistency of the
perturbations at first order. Here $\Box _{\eta}$ is the D'Alambertian
operator compatible with the flat metric $\eta _{a b }$. 

The following gauge condition will be imposed in order to simplify the 
linearized equations: 
\begin{equation}  \label{radgaugescal0}
\partial^c \tilde \gamma_{cb} = 0 \,\,\,\,.
\end{equation}
As it is shown below, for the choice of $\kappa$ given by 
Eq. (\ref{kappa}), this gauge condition corresponds to the 
linear limit of the pseudo-harmonic gauge Eq. (\ref{STPHG}) for $\Theta=1$.

 Then, from Eq. (\ref{einstlin}) the resulting wave equation is 
\begin{equation}  \label{wavetilgamma}
\Box_\eta \tilde \gamma_{ab} = - 16\pi G_0 \tilde T_{ab} - 2\kappa
\left( \partial^2_{ab}\tilde\phi -\eta_{ab}\Box_\eta \tilde\phi
\right) \,\,\,\,.
\end{equation}

The linear approximation of the effective energy-momentum tensor Eq. (\ref
{effTmunu}) and the Eq. (\ref{KG}) turn to be 
\begin{eqnarray}\label{Tmunulin}
& & \tilde T_{ab} = \frac{\tilde T_{ab}^{{\rm matt}}}{8\pi G_0 f_0} + 
\frac{f^{\prime}_0}{8 \pi G_0 f_0} \left( \partial^2_{ab}\tilde\phi
-\eta_{ab}\Box_\eta \tilde\phi \right) \,\,\,, \\
&& \Box_\eta \tilde \phi - m_0^2 \tilde \phi = 4\pi \alpha 
\frac{f^{\prime}_0}{f_0} \tilde T_{{\rm matt}} \,\,\,\,, \\
\label{msq}
&& m_0^2 := \frac{V^{\prime\prime}_0}{1 + \frac{ 3{f^{\prime}_0}^2}{2
f_0}} \,\,\,\,, \\
\label{alp}
&& \alpha := \frac{1}{8\pi \left(1 + \frac{ 3{f^{\prime}_0}^2}{2 f_0}\right)} 
\,\,\,\,,
\end{eqnarray}
where the following conditions were used 
\begin{equation}  \label{V0cond}
V(\phi_0) = 0 = V^{\prime}_0 = 0 \,\,\,\,,
\end{equation}
resulting from the consistency at first order of the linearized Einstein and 
scalar-field equations and assuming that $T_{{\rm matt}\,\,\,ab}^0 = 0$.
In this way the wave equation (\ref{wavetilgamma}) becomes 
\begin{equation}  \label{wavetilgamma2}
\Box_\eta \tilde \gamma_{ab} = - \frac{2}{f_0} 
\tilde T_{ab}^{{\rm matt}} - 2\left(\kappa + \frac{f^{\prime}_0}{f_0}\right)
\left( \partial^2_{ab}\tilde\phi -\eta_{ab}\Box_\eta \tilde\phi
\right) \,\,\,\,.
\end{equation}
Note that at this order of approximations the application of the ordinary divergence in Eq. 
(\ref{wavetilgamma2}) and the use of the gauge Eq. (\ref
{radgaugescal0}) leads to the energy-conservation of the matter
perturbations: $\partial^a \tilde T_{ab}^{{\rm matt}}=0$.

The convenient choice for $\kappa$,
\begin{equation}
\label{kappa}
\kappa =-\frac{f_{0}^{\prime }}{f_{0}}\,\,\,\,,  
\end{equation}
allows to simplify Eq. (\ref{wavetilgamma2}) considerably. In summary, one deals with the following wave equations for the
gravitational and scalar modes \footnote{It is not a surprise to recognize in 
$\tilde{\gamma}_{a b }$, the conformal metric perturbation $\bar{\gamma}_{a b }^*/F_0$ which is 
defined from $g_{ab}^* \approx F_0\left[\eta_{ab} + \epsilon \left(\gamma_{ab} + \frac{F_0^\prime}{F_0}\tilde\phi \eta_{ab}\right)\right]$
as $\bar{\gamma}_{a b }^*:= \gamma _{a b }^*-\frac{1}{2}\eta _{a b}\gamma^*$.}: 
\begin{eqnarray}
&&\Box _{\eta }\tilde{\gamma}_{a b }=-\frac{2}{f_{0}}\tilde{T}_{a b }^{{\rm matt}}\,\,\,\,,  \label{wavetilgamma3} \\
&&\partial ^{a }\tilde{\gamma}_{a b }=0\,\,\,\,, \\
\label{KGlin}
&&\Box _{\eta }\tilde{\phi}-m_{0}^{2}\tilde{\phi}=4\pi \alpha \frac{f_{0}^{\prime }}{f_{0}}\tilde{T}_{{\rm matt}}\,\,\,\,,
\end{eqnarray}
with the constants $m_{0}^{2},\alpha$ and $\kappa$ given by Eqs. 
(\ref{msq}), (\ref{alp}) and (\ref{kappa}) respectively. 
The analysis of propagation of gravitational
and scalar waves will be not pursued here, this has been done elsewhere 
(see Refs. \cite{scalwaves} ).

\subsection{Linear limit of 3+1 equations}
In order to linearize the 3+1 
set of equations (\ref{K_ij}), (\ref{Pi}), (\ref{evQ2})
and (\ref{CEHfSST})$-$(\ref{evPi1}), once again assume first order deviations of 
the Lorentz metric $\eta _{a b }$ as follows 
\begin{eqnarray}
N &\approx &1+\epsilon \tilde{N}\,\,\,\,\,, \\
N_{i} &\approx &\epsilon \tilde{N}_{i}\,\,\,\,\,, \\
h_{ij} &\approx &\delta _{ij}+\epsilon \tilde{h}_{ij}\,\,\,\,\,,
\end{eqnarray}
where $\epsilon \ll 1$.

Note that at first order the inverse of the 3-metric is given by
\begin{equation}
\label{h^ijaprox}
h^{ij}\approx \delta _{ij}-\epsilon \tilde{h}_{ij} \,\,\,,
\end{equation}
such that the condition $h_{il}h^{lj}=\delta _{\,\,\,j}^{i}$ holds.
Therefore, covariant and contravariant components of tensorial quantities
having no zero order terms are identical to each other at first order. For
instance $N^{i}=h^{il}N_{l}\approx \epsilon \tilde{N}_{i}$. In order to
compare with the 4+0 (covariant) linear approximation 
\begin{equation}
g_{a b }\approx \eta _{a b }+\epsilon \gamma _{a b }\,\,\,\,,
\end{equation}
it turns out that 
\begin{eqnarray}
\gamma _{00} &=&-2\tilde{N}\,\,\,\,,  \label{Newtpot3p1} \\
\gamma _{0i} &=&-\tilde{N}_{i}\,\,\,\,, \\
\gamma _{ij} &=&\tilde{h}_{ij}\,\,\,\,.
\end{eqnarray}

The 3-Christoffel symbols turn to be 
\begin{eqnarray}
^{3}\,\!\Gamma _{jk}^{i} &\approx &\epsilon \,\,^{3}\,\!\tilde{\Gamma}%
_{jk}^{i}\,\,\,\,, \\
^{3}\,\!\tilde{\Gamma}_{jk}^{i} &:=&\frac{1}{2}\left( -\partial _{i}\tilde{h}%
_{jk}+\partial _{k}\tilde{h}_{ij}+\partial _{j}\tilde{h}_{ki}\right)
\,\,\,\,.
\end{eqnarray}
Therefore 3-covariant derivatives of 3-tensors having no zero order terms
become at first order ordinary derivatives. For instance $^{3}\nabla
_{j}N_{i}\approx \epsilon \partial _{j}\tilde{N}_{i}$. Then Eq. (\ref{K_ij})
leads to 
\begin{eqnarray}
K_{ij} &\approx &\epsilon \tilde{K}_{ij}\,\,\,\,, \\
\label{linK_ij}
\tilde{K}_{ij} &:= &-\frac{1}{2}\left( \partial _{t}\tilde{h}_{ij}+2\partial
_{(i}\tilde{N}_{j)}\right) \,\,\,\,,
\end{eqnarray}
with 
\begin{eqnarray}
K_{\,\,\,j}^{i} &\approx &\epsilon \tilde{K}_{\,\,\,j}^{i}\approx \epsilon 
\tilde{K}_{ij}\,\,\,\,,\\
K &:= &K_{\,\,\,l}^{l}\approx \epsilon \tilde{K}\,\,\,\,, \\
\label{Klin}
\tilde{K} &:=& -\frac{1}{2}\left( \partial _{t}\tilde{h}+2\partial _{l}\tilde{%
N}^{l}\right) \,\,\,\,,
\end{eqnarray}
where $\tilde{h}:=\tilde{h}_{ll}$ is the trace of the 3-metric perturbation.
Using the definition of the 3-Riemann tensor in terms of the 3-Christoffel
symbols it is easy to obtain the linearized approximation for the 3-Ricci
tensor and the 3-curvature respectively: 
\begin{eqnarray}
^{3}R_{ij} &\approx &\epsilon \,\,^{3}\tilde{R}_{ij}\,\,\,\,, \\
^{3}\tilde{R}_{ij} &:= &\frac{1}{2}\left( 
2\partial_{l(j}^{2}\tilde{h}_{i)l}^{}-\,^{3}\tilde{\Delta}\tilde{h}_{ij}-\partial _{ij}^{2}\tilde{h}\right)
\,\,\,\,, \\
^{3}R &\approx &\epsilon \,\,^{3}\tilde{R}\,\,\,\,, \\
^{3}\tilde{R} &:= &\partial _{kl}^{2}\tilde{h}_{kl}-\,^{3}\tilde{\Delta}%
\tilde{h}\,\,\,\,,
\end{eqnarray}
where $^{3}\tilde{\Delta}:=\partial _{ll}^{2}$ stands for the Euclidean
3-Laplacian.

Concerning the scalar field and the matter, in addition to 
the approximations given by (\ref{scalaproxi})$-$(\ref{scalaproxf}), the following is also to be assumed
\begin{eqnarray}
Q_i &\approx & \epsilon \tilde Q_i \,\,\,\,,\\
\label{Qlin}
\tilde Q_i &:=& \partial_i\tilde \phi\,\,\,\,,\\
\Pi &\approx &\epsilon \tilde \Pi \,\,\,\,,\\
\label{Pilin}
 \tilde \Pi &:=& \partial_t \tilde \phi \,\,\,\,,\\
E^{\rm matt}&\approx & E_0^{\rm matt} + \epsilon \tilde E^{\rm matt} \,\,\,\,, \\
J_i^{\rm matt} &\approx & J^{0\,\,{\rm matt}}_i + 
\epsilon \tilde J_i^{\rm matt} \,\,\,\,, \\
S_{ij}^{\rm matt} &\approx & S_{ij}^{0\,\,{\rm matt}} + 
\epsilon \tilde S_{ij}^{\rm matt} \,\,\,\,,
\end{eqnarray}
where the knot indicates quantities at zero order.

The Hamiltonian constraint Eq. (\ref{CEHfSST})
 when linearized reads
\begin{equation} 
\label{CEHflin0}
^3 \tilde R - \frac{2f^\prime_0}{f_0} \partial_l {\tilde Q}^l 
  = \frac{2 \tilde{E}_{\rm matt}}{f_0}  \,\,\,,
\end{equation}
or explicitly 
\begin{equation}  
\label{CEHflin}
\partial^2_{kl}\tilde h_{kl} - \,^3\tilde \Delta \tilde h 
- \frac{2f^\prime_0}{f_0} \partial_l {\tilde Q}^l  = 
\frac{2\tilde{E}_{\rm matt}}{f_{0}}
 \,\,\,.
\end{equation}

The linear limit of the momentum constraint Eq. (\ref{CEMfSST}) is
\begin{equation}  
\label{CEMflin}
\partial_l \tilde K_{\,\,\,\,\,i}^{l} - \,\partial_i \tilde K 
+\frac{f^\prime_0}{f_0}\partial_i \tilde{\Pi}
= \frac{{\tilde J}_i^{\rm matt}}{f_0} 
\end{equation}

Finally the dynamic Einstein Eqs. (\ref{EDEfSST}) when linearized read 
\begin{eqnarray}  
\label{EDEflin}
&&
\partial_t \tilde K_{\,\,\,j}^i + \partial^i\,\partial_j \tilde N - \,^3 
\tilde R_{\,\,\,j}^i 
+ \frac{f^\prime_0}{f_0}\partial^i {\tilde Q}_j =
-\frac{1}{2f_0\left(1 + \frac{3{f^\prime_0}^2}{2f_0}\right)}
\left\{ 2 \tilde{S}_{{\rm matt}\,\,\,j}^i \left(1 + \frac{3{f^\prime_0}^2}{2f_0} \right)
\right. \nonumber \\
&& \left.
+ \delta_{\,\,\,j}^i \left[ f^\prime_0 V^{\prime\prime}_0 \tilde{\phi} 
- \left(\rule{0mm}{0.4cm} \tilde{S}_{\rm matt}- 
\tilde{E}_{\rm matt}\right) \left(1 + \frac{{f^\prime_0}^2}{f_0} \right)
\right]\right\}\,. 
\end{eqnarray}

The linear limit of the evolution Eq. (\ref{EDKSTT}) reads 
\begin{equation}  \label{EDK2lin}
\partial_t \tilde K + \,^3\tilde \Delta \tilde N 
- \frac{f^\prime_0}{f_0}\partial_l \tilde{Q}^l  
 = 
\frac{1}{2f_0\left(1 + \frac{3{f^\prime_0}^2}{2f_0}\right)}
\left\{ \tilde{S}_{\rm matt} + \tilde{E}_{\rm matt} 
\left(1 + \frac{3{f^\prime_0}^2}{f_0}\right) - 3f^\prime_0 
V^{\prime\prime}_0\tilde{\phi} \right\}
 \,\,\,.
\end{equation}

Finally, the linear limit of the Eqs. (\ref{evQ2}) and (\ref{evPi1}) takes respectively 
the following form
\begin{equation}
\label{linevQ}
\partial_t \tilde{Q}_i= \partial_i \tilde{\Pi}\,\,\,\,,
\end{equation}

\begin{equation}
\label{linevPi1}
\partial_t \tilde{\Pi}  - \partial_j \tilde{Q}^j  
= - \frac{ f_0 V^{\prime\prime}_0 \tilde{\phi} 
 + \frac{1}{2}f^\prime_0 \tilde{T}_{{\rm matt}} }
{f_0\left(1 + \frac{3{f^\prime_0}^2}{2f_0}\right) } \,\,\,.
\end{equation}
 Equations (\ref{Qlin}), (\ref{Pilin})  and  (\ref{linevPi1}) are indeed equivalent to Eq. (\ref{KGlin}). One reminds $\tilde{T}_{{\rm matt}}= \tilde{S}_{{\rm matt}}-\tilde{E}_{{\rm matt}}$.

It is to be emphasized that in the above linearization 
the self-consistency of the 3+1 equations up to first order imply that the
zero order source fields must vanish identically: 
\begin{eqnarray}
E_0^{\rm matt} = 0 = J^{0\,\,{\rm matt}}_i = 
S_{ij}^{0\,\,{\rm matt}}= V^\prime_0= V_0 \,\,\,\,.
\end{eqnarray}

The above equations (\ref{CEHflin})$-$(\ref{EDEflin})
are the 3+1 decomposition of the 4+0 equations (\ref{einstlin}).

It is to note that the 3+1 splitting of the perturbations (\ref{tilgamma}) is given by
\begin{eqnarray}
\tilde{\gamma}_{00} &=&\frac{1}{2}\left( \tilde{h}-2\kappa \tilde{\phi}-2%
\tilde{N}\right) \,\,\,\,, \\
\tilde{\gamma}_{0i} &=&-\tilde{N}_{i}\,\,\,\,, \\
 \label{tilgammaij}
\tilde{\gamma}_{ij} &=&\tilde{h}_{ij}-\frac{1}{2}\delta _{ij}\left( \tilde{h}%
-2\kappa \tilde{\phi}+2\tilde{N}\right) \,\,\,\,, \\
\tilde{\gamma} &=&-\left( \tilde{h}+2\tilde{N}\right) +4\kappa \tilde{\phi}%
\,\,\,\,.
\end{eqnarray}

Now, assuming the PHG given by Eqs. (\ref{STTHlapse}) and (\ref{STTHshift}) and taking 
$\kappa= - \Theta {\cal F}^\prime_0=  - \Theta f^\prime_0/f_0 $, the linear limit reads
\begin{eqnarray} 
\partial _{t}\tilde{N} + \kappa \tilde \Pi +\tilde{K} =0 \,\,\,\,,
\label{radgauge0scal} \\
\label{radgaugeiscal}
\partial_t \tilde N_i + \partial_j \tilde h_{ij} -\partial_i \tilde N
-\frac{1}{2}\partial_i\tilde h + \kappa \tilde Q_i = 0 \,\,\,\,,
\end{eqnarray}
where the following linear approximations were used $V^i= 
 \frac{1}{\sqrt{h}}\partial_j\left(\sqrt{h} h^{ij}\right) \approx \epsilon \tilde V_i$
with $\tilde V_i= -\left(\partial_j \tilde h_{ij} -\frac{1}{2}\partial_i\tilde h\right)$. 
This result is obtained using Eq. (\ref{h^ijaprox}) and also from the fact that at first order 
$h={\rm det}h_{ij}\approx 1+\epsilon \tilde h$. 
Both Eqs. (\ref{radgauge0scal}) and (\ref{radgaugeiscal}) amounts to the 3+1 decomposition 
of the  Lorentz-De Donder gauge Eq. (\ref{radgaugescal0}).

From Eqs. (\ref{radgauge0scal}), (\ref{EDK2lin}), and (\ref{linevPi1})
 one obtains a wave equation for $\tilde{N}$: 
\begin{equation}
\Box_\eta \tilde{N} -\left(\kappa + \frac{f^\prime_0}{f_0}\right)\partial_l \tilde{Q}^l
=\frac{1}{2f_0\left(1 + \frac{3{f^\prime_0}^2}{2f_0}\right)}
\left\{ \tilde{S}_{\rm matt}\left(1-\kappa f^\prime_0\right) 
+ \tilde{E}_{\rm matt} 
\left(1 + \frac{3{f^\prime_0}^2}{f_0}+ \kappa f^\prime_0 \right) 
- 2 f_0 V^{\prime\prime}_0\tilde{\phi}
\left(\kappa  + \frac{3f^\prime_0}{2f_0}\right) \right\}
 \,\,\,.
\label{waveNscal}
\end{equation}
where $\Box_\eta:= -\partial _{tt}^{2} +\,^{3}\tilde{\Delta}$. 

On the other hand, the combination of Eqs. (\ref{CEMflin}) and (\ref{linK_ij}) with 
Eqs. (\ref{radgaugeiscal}) and (\ref{radgauge0scal}) 
(noting $\partial_t {\tilde Q}_i= \partial_i {\tilde \Pi}$ which follows from the integrability 
condition $\partial_{ti}^2{\tilde \phi}=\partial_{it}^2{\tilde \phi}$), results in a wave equation for $\tilde{N}^{i}$: 
\begin{equation}
\Box_\eta \tilde N^i -2\left( \kappa + \frac{f^{\prime}_0}{f_0}
\right) \partial_i\tilde\Pi = -\frac{2 \tilde J^{i}_{{\rm matt}}}{f_0}    \,\,\,\,.
\label{waveNiscal}
\end{equation}
Finally, the linearized dynamic Einstein Eqs. (\ref{EDEflin}) together with
Eqs. (\ref{linK_ij}) and (\ref{radgaugeiscal}) lead to a wave equation for $\tilde{h}_{ij}$ as follows,
\begin{eqnarray}
&&
\frac{1}{2} \Box_\eta \tilde{h}_{ij} + \left(\kappa + \frac{f^\prime_0}{f_0}\right)\partial_i 
\tilde{Q}_j
=
-\frac{1}{2f_0\left(1 + \frac{3{f^\prime_0}^2}{2f_0}\right)}
\left\{ 2 \tilde{S}^{\rm matt}_{ij} \left(1 + \frac{3{f^\prime_0}^2}{2f_0} \right)
\right. \nonumber \\
&& \left.
+ \delta_{ij} \left[ f^\prime_0 V^{\prime\prime}_0 \tilde{\phi} 
- \left(\rule{0mm}{0.4cm} \tilde{S}_{\rm matt}- 
\tilde{E}_{\rm matt}\right) \left(1 + \frac{{f^\prime_0}^2}{f_0} \right)
\right]\right\}\,\,\,\,. 
\label{wavehijscal}
\end{eqnarray}
From Eq. (\ref{wavehijscal}) and tracing one obtains 
\begin{eqnarray}
&&
\frac{1}{2} \Box_\eta\tilde{h}+ \left(\kappa + \frac{f^\prime_0}{f_0}\right)\partial_l 
\tilde{Q}^l
=
-\frac{1}{2f_0\left(1 + \frac{3{f^\prime_0}^2}{2f_0}\right)}
\left\{ -\tilde{S}_{\rm matt} + 3f^\prime_0 V^{\prime\prime}_0 \tilde{\phi} 
+ 3\tilde{E}_{\rm matt} \left(1 + \frac{{f^\prime_0}^2}{f_0} \right)
\right\}\,\,\,\,. 
\label{wavehscal}
\end{eqnarray}
Therefore combining Eqs. (\ref{waveNscal}) and (\ref{wavehscal}) yields
\begin{equation}
\label{wavebargamma00scal}
\Box_\eta \,\left( \frac{\tilde{h}}{2}
-\tilde{N}\right) + 2 \left(\kappa + \frac{f^\prime_0}{f_0}\right)\partial_l 
\tilde{Q}^l
 =
-\frac{1}{2f_0\left(1 + \frac{3{f^\prime_0}^2}{2f_0}\right)}
\left\{ -\kappa f^\prime_0 \left(\rule{0mm}{0.4cm} \tilde{S}_{\rm matt}- 
\tilde{E}_{\rm matt}\right) - 2 \kappa f_0 V^{\prime\prime}_0 \tilde{\phi} 
+ 4\tilde{E}_{\rm matt} \left(1 + \frac{3{f^\prime_0}^2}{2f_0} \right)
\right\}\,\,\,\,. 
\end{equation}
Finally one can further combine the wave equation (\ref{linevPi1}) and (\ref{wavebargamma00scal}) 
to find 
\begin{equation}
\label{wavebargamma00scal2}
\Box_\eta\,\left( \frac{\tilde{h}}{2}
-\tilde{N} -\kappa \tilde{\phi} \right) + 2 \left(\kappa + \frac{f^\prime_0}{f_0}\right)\partial_l 
\tilde{Q}^l = -\frac{2 \tilde{E}_{\rm matt}}{f_0}\,\,\,\,. 
\end{equation}
Perhaps a more straightforward way of obtaining this wave equation is 
by applying $\partial_t$ and $\partial^i$ to Eqs. (\ref{radgauge0scal}) and (\ref{radgaugeiscal}) 
respectively and then adding the resulting equations followed by the use of Eqs. (\ref{CEHflin}) and (\ref{Klin}).

On the other hand, the combination as given by Eq. (\ref{tilgammaij}) provides the wave equation 
\begin{equation}   
\label{wavetilgammaij}
\Box_\eta\,\tilde \gamma_{ij} + 
2 \left(\kappa + \frac{f^\prime_0}{f_0}\right)\partial_i \tilde{Q}_j
= -\frac{2 \tilde{S}^{\rm matt}_{ij}}{f_0} 
+\frac{\delta_{ij}\left(\kappa + \frac{f^\prime_0}{f_0}\right)}{f_0\left(1 + \frac{3{f^\prime_0}^2}{2f_0}\right)}
\left[\rule{0mm}{0.5cm} f^\prime_0 \left(\rule{0mm}{0.4cm} \tilde{S}_{\rm matt}- 
\tilde{E}_{\rm matt}\right) + 2 f_0 V^{\prime\prime}_0 \tilde{\phi} 
\right]\,\,\,\,. 
\end{equation}

Therefore Eqs. (\ref{waveNiscal}), (\ref{wavebargamma00scal2}), and 
(\ref{wavetilgammaij}) are equivalent to Eq. (\ref{wavetilgamma2}). In such 
equations one can employ the choice $\kappa =-f_{0}^{\prime}/f_{0}$ (which corresponds to $\Theta=1$) 
to simplify the expressions as follows
\begin{eqnarray}
\label{wavebargamma00scal2gauge}
\Box_\eta\,\left( \frac{\tilde{h}}{2}
-\tilde{N} +\frac{f^\prime_0}{f_0} \tilde{\phi} \right) &=& -\frac{2 \tilde{E}_{\rm matt}}{f_0}\,\,\,\,,\\
\label{waveNiscalgauge}
\Box_\eta \tilde N^i &=& -\frac{2 \tilde J^{i}_{{\rm matt}}}{f_0} \,\,\,\,,\\
\label{wavetilgammaijgauge}
\Box_\eta\,\tilde \gamma_{ij} 
& = & -\frac{2 \tilde{S}^{\rm matt}_{ij}}{f_0} \,\,\,\,. 
\end{eqnarray}
These equations are equivalent to Eq. (\ref{wavetilgamma3}).

In fact, with $\kappa =-f_{0}^{\prime}/f_{0}$, another useful wave equation is obtained from Eqs. (\ref{waveNscal}) and 
(\ref{KGlin}), 
\begin{equation}
\label{waveNphi}
\Box_\eta\,\left(\tilde{N} +\frac{f^\prime_0}{2f_0} \tilde{\phi} \right)= 
\frac{\tilde{S}_{\rm matt} + \tilde{E}_{\rm matt}  }{2f_0} \,\,\,\,.
\end{equation}

\subsection{The Newtonian approximation}

\smallskip 
Since the complete equivalence between the linear limits of the 
covariant field equations and the 3+1 equations has been shown, 
the Newtonian approximation (slow varying sources and small pressures) will be 
considered only from the linear limit of the 3+1 approach. 
In order to do so, one takes $\tilde{E}_{{\rm matt}}=\tilde{\rho}$
, $\tilde{S}^{{\rm matt}} \ll \tilde{\rho},\tilde{J}_{{\rm matt}}^{i} \ll \tilde{
\rho}c$ and neglect the time dependence of the perturbations. 
From Eqs. (\ref{waveNiscalgauge}) and (\ref{wavetilgammaijgauge}) 
where one assumes $\tilde{S}^{{\rm matt}}_{ij}\approx 0 \approx \tilde{J}_{{\rm matt}}^{i}
$ and taking into 
account that perturbations are well behaved at infinity, one concludes that the unique solutions are 
$N^i= {\rm const.}$ and $\tilde \gamma_{ij}={\rm const.}$ 
Such constants can be gauged out and so ${\tilde N}^i= 0 = \tilde \gamma_{ij} $. 

On the other hand, Eqs. (\ref{KGlin}) and (\ref{waveNphi}) lead respectively to 
\begin{equation}
\,^{3}\tilde{\Delta} \tilde{\phi} - 
\frac{ V^{\prime\prime}_0 \tilde{\phi}}
{1 + \frac{3{f^\prime_0}^2}{2f_0} }
= -\frac{f^\prime_0 \tilde{\rho} }
{2f_0\left(1 + \frac{3{f^\prime_0}^2}{2f_0}\right) }
 \,\,\,,
\label{weakphi}
\end{equation}
\begin{equation}
\,^{3}\tilde{\Delta}\left( \tilde{N}+\frac{f_{0}^{\prime }}{2f_{0}}\,\tilde{
\phi}\,\right) =\frac{\tilde{\rho} }{2 f_0}\,\,\,\,.  \label{laplaceNscal}
\end{equation}
The Newtonian approximation in the geodesic equation for test particles leads to the 
identification of $\tilde N$ with the Newtonian potential $\Phi_N$. Hence, from 
Eqs. (\ref{weakphi}) and (\ref{laplaceNscal}) one finds the solution
\begin{widetext}
\begin{equation}
\Phi _{N}= -\frac{1}{8\pi f_{0}}\int \frac{\tilde{\rho}(\vec{x}^{\prime })}{|
\vec{x}-\vec{x}^{\prime }|}d^{3}x^{\prime }\,\,\,\, -\frac{\alpha }{2}\left( 
\frac{f_{0}^{\prime }}{f_{0}}\right) ^{2}\int \frac{\tilde{\rho}(\vec{x}
^{\prime })e^{-m_{0}|\vec{x}-\vec{x}^{\prime }|}}{|\vec{x}-\vec{x}^{\prime }|
}d^{3}x^{\prime }\,\,\,\,\,\,+\,\,\,\,\,\,{\rm B.C.}  \label{Newtpot}
\end{equation}
\end{widetext}
where $m_0$ and $\alpha$ are given by Eqs. (\ref{msq}) and (\ref{alp}), respectively.

For instance, in the exterior of an extended body with mass-density $\tilde{\rho}$ 
with compact support, the Newtonian potential is given by 
\begin{equation}
\Phi _{N}=-\frac{G_{0}M}{F_{0}r}\left[ 1+\frac{\alpha}{2G_{0}}\frac{%
(F_{0}^{\prime })^{2}}{F_{0}}e^{-m_{0}r}\right] \,\,\,\,,
\end{equation}
where $r= |\vec{x}|$, and $f_0=F_0/(8\pi G_0)$ was restored. 
In the above expression $M$ corresponds to the gravitational mass of the body given by the 
volume integral of the mass-density $\tilde{\rho}$ 
\footnote{In the strong-field regime (e.g., neutron stars) 
and for instance in the static and spherically symmetric case, 
the asymptotic behavior for 
the lapse and the scalar-field are $N\sim 1-G_0 M_{\rm ADM}/r$ 
and $\phi \sim \omega_\phi/r$ (for $m_0\equiv 0$ and $\phi_0\equiv 0$), where $\omega_\phi$ 
is the {\it scalar charge} \cite{salgado2} and $M_{\rm ADM}$ is the ADM mass.}.

It is customary to express $F_{0}^{\prime }$ in
terms of the effective Brans-Dicke parameter Eq. (\ref{BDpar}) 
\begin{equation}
\omega _{{\rm BD}}^{0}=\frac{8\pi G_{0}F_{0}}{(F_{0}^{\prime })^{2}}\,\,\,\,.
\end{equation}
Then, 
\begin{equation}
\alpha =\frac{\omega _{{\rm BD}}^{0}}{4\pi \left( 3+2\omega _{{\rm BD}%
}^{0}\right) }\,\,\,\,,\,\,\,\,
m_{0}^{2} =\frac{2\omega _{{\rm BD}}^{0}V_{0}^{\prime \prime }}{3+2\omega
_{{\rm BD}}^{0}}\,\,\,\,,
\end{equation}
therefore,
\begin{equation}
\Phi _{N}=-\frac{G_{0}M}{F_{0}r}\left[ 1+\frac{e^{-m_{0}r}}{3+2\omega _{%
{\rm BD}}^{0}}\right] \,\,\,\,.  \label{Newtpot2}
\end{equation}

Different experiments bound the parameters (like $\omega _{{\rm BD}}$ and $V_{0}^{\prime \prime }$) 
of theories that induce modifications of the Newtonian force law $\sim 1/r^2$ (see for instance Refs. \cite{adelberger,Baldi,lageos,cassini,jimenez}).

Now, for the massless case $m_{0}=0$, the Newtonian potential reads
\begin{equation}
\Phi _{N}=-\frac{G_{\rm cav} M}{r} \,\,\,\,,  \label{Newtpot3}
\end{equation}
where 
\begin{equation}
G_{\rm cav}:=\frac{2G_{0}}{F_{0}}\left( \frac{2+\omega _{{\rm BD}}^{0}}{3+2\omega _{%
{\rm BD}}^{0}}\right) \,\,\,\,,
\end{equation}
is the {\it dressed} gravitational constant measured by a Cavendish experiment
\cite{damour1,riazuelo2}.

\section{3+1 Decomposition of Differential Operators}
\label{sec:appendix}
Let us consider the D'Alambertian of any number $A$ of functions
$\psi^A(x^a)$ with their corresponding source terms:
\begin{equation}
\Box \psi^A = S^{\psi^A}\,\,\,.
\end{equation}
Now,
\begin{equation}
\Box \psi^A = \frac{1}{\sqrt{-g}}\partial_c\left[\sqrt{-g} g^{cd}\partial_d \psi^A\right]\,\,\,\,. 
\end{equation}
Using $g^{ab}= h^{ab} - n^a n^b$ and the fact that 
$g:= {\rm det}g_{ab}= -N^2 h$ where $h:= {\rm det}h_{ij}$, then
\begin{equation}
\Box \psi^A = \frac{1}{N\sqrt{h}}
\partial_c\left[N\sqrt{h}\,h^{cd}\partial_d \psi^A\right]
- \frac{1}{N\sqrt{h}}
\partial_c\left[N\sqrt{h}\, n^c n^d \partial_d \psi^A\right]\,\,\,\,.
\end{equation}
Hence,
\begin{equation}
\Box \psi^A = \,^3\Delta \psi^A + a^c \nabla_c \psi^A + K n^c\nabla _c \psi^A
- n^c \nabla _c \left( n^d \nabla_d \psi^A\right)\,\,\,.
 \end{equation}
where one reminds $^3\Delta$ is the Laplacian compatible with the 3-metric $h_{ij}$ and 
$a^c= n^d \nabla_d n^c\equiv D^c[{\rm ln}N]$ is the 4-acceleration 
of the observer with 4-velocity $n^a$. Here it was used the identity $K= - \nabla_c n^c$. 
In order to obtain a system of first order equations one can further define 
as in Eqs. (\ref{Q}) and (\ref{Pidef}) the following quantities
\begin{eqnarray}
\label{QA}
Q^{a A}&:=& D^a \psi^A \,\,\,\,,\\
\label{PiA}
\Pi^A &:=& {\cal L}_{\mbox{\boldmath{$n$}}} \psi^A= n^c \nabla_c \psi^A\,\,\,\,.
\end{eqnarray}
Collecting the above results one obtains
\begin{equation}
\label{3+1Box}
{\cal L}_{\mbox{\boldmath{$n$}}} \Pi^A - a_c Q^{c A} - D_c Q^{c A}- \Pi^A 
K= -  S^{\psi^A}\,\,\,.
\end{equation}
A simple application of the above results is the case when 
$\psi^A\equiv x^A$  (with $x^A= (t,x^i)$ as in Sec. IV). It turns then $\Pi^A\equiv n^A=
\frac{1}{N}(1,N^i)$ and $Q^{\,\,A}_c= h^{\,\,d}_c\delta^{\,\,A}_d
\equiv h_c^{\,\,A}$ [where it is to note that $Q^{\,\,0}_c\equiv 0$ and 
 $D_c Q^{c i}= \frac{1}{\sqrt{h}} 
\partial_j\left(\sqrt{h} h^{ij}\right) = V^i$]. The above 
Eq. (\ref{3+1Box}) with a suitable choice of $S^{\psi^A}$ 
is then equivalent to Eq. (\ref{STTBMev}) which 
provides an evolution equation for 
the lapse and for the shift. For instance, taking $S^{\psi^A}\equiv 0$ 
in Eq. (\ref{3+1Box}) allows one to recover Eqs. (\ref{STTHlapse}) and (\ref{STTHshift}) 
(for $\Theta=0$) which are equivalent to the 
{\it harmonic coordinate condition} Eq. (\ref{HG}). 
 
On the other hand, one can take $\Box \psi^A= S^{\psi^A}$ with 
$S^{\psi^A}= \Upsilon_{\psi^A} n^c n^d \nabla_c\nabla_d \psi^A -
\Theta\nabla^A {\cal F}$ (no sum over index $A$). Now, using $n^c n^d= h^{cd} - 
g^{cd}$, one 
obtains
\begin{equation}
n^c n^d \nabla_c\nabla_d \psi^A= 
h^{cd} \nabla_c\nabla_d \psi^A - \Box \psi^A \,\,\,.
\end{equation}
Using the orthogonal decomposition (see Sec. III) 
$\nabla_a \psi^A= D_a \psi^A - n_a n_c \nabla^c \psi^A
= Q_a^A - n_a \Pi^A$ it yields
\begin{equation}
n^c n^d \nabla_c\nabla_d \psi^A= D_c Q^{c A} + K \Pi^A 
- \Box \psi^A\,\,\,,
\end{equation}
where one used $h^{cd}\nabla_c n_d= \nabla_c n^d= -K$ and 
$h^{cd} n_d\equiv 0$. 
In this way the equation $\Box \psi^A= S^{\psi^A}$ becomes
\begin{equation}
\Box \psi^A= \frac{1}{1+\Upsilon_{\psi^A}}\left[ 
\Upsilon_{\psi^A} \left(D_c Q^{c A} + K \Pi^A\right) -
\Theta\nabla^A {\cal F}\right]
\,\,\,.
\end{equation}
Finally, $-\Box \psi^A$ is given by the l.h.s of Eq. (\ref{3+1Box}) which leads to 
\begin{equation}
{\cal L}_{\mbox{\boldmath{$n$}}} \Pi^A - a_c Q^{c A} - D_c Q^{c A}- \Pi^A 
K= -\frac{1}{1+\Upsilon_{\psi^A}}\left[ 
\Upsilon_{\psi^A} \left(D_c Q^{c A} + K \Pi^A\right) -
\Theta\nabla^A {\cal F} \right]
\,\,\,,
\end{equation}
which simplifies to give
\begin{equation}
\label{3+1BoxS}
{\cal L}_{\mbox{\boldmath{$n$}}} \Pi^A - a_c Q^{c A}
= \frac{1}{1+\Upsilon_{\psi^A}}\left(D_c Q^{c A} + K \Pi^A + 
\Theta\nabla^A {\cal F} \right)\,\,\,.
\end{equation}
As before, with $\psi^A= x^A= (t,x^i)$ and $\Upsilon_{\psi^A}=\Upsilon_{q_A}$, and 
using Eqs. (\ref{Q}) and (\ref{Pi}) (which leads to $\Pi^A= n^A=
\frac{1}{N}(1,N^i)$ and $Q^A_c= h^{\,\,d}_c\delta^A_d
\equiv h_c^A$), from the above equation (\ref{3+1BoxS}) one recovers 
Eq. (\ref{STTBMev}) for $B=0$, $A=1$. Following the above examples and 
taking for $S^{\psi^A}$ the 
r.h.s of Eq. (\ref{STTBM}) one can then generalize the expression of 
Eq. (\ref{3+1BoxS}) so as to obtain Eqs. (\ref{STTlapse}) and (\ref{STTlapseshift}).

\section{Self-consistency of 3+1 Field Equations}
As mentioned in Sec. I, one of the main properties of the STT is that they preserved the weak equivalence principle in the sense that 
the conservation of the effective energy-momentum tensor (\ref{effTmunu})
\begin{equation}
\label{consTeff}
\nabla _{b }T^{ab}=0\,\,\,\,,
\end{equation}
together with the field equations (\ref{Einst}) and (\ref{KGo}) lead to the conservation of the energy-momentum tensor of matter alone
\begin{equation}
\nabla _{b }T_{\rm matt}^{ab}=0\,\,\,\,.
\end{equation}
The same result can be expressed in terms of the 3+1 decomposition of the field equations as performed in Sec. III. However, it is not 
at all evident that the same result will turn out if one has committed any mistake (one as ``innocent'' as an error sign) during the 
process of the derivation of the 3+1 field equations of STT. In order to carry out such a check one requires to express the conservation 
equation (\ref{consTeff}) in 3+1 form. Without showing any mathematical details, the projection of  Eq. (\ref{consTeff}) on $\Sigma_t$ 
and using Eq. (\ref{Tort}) leads to the equation of conservation of momentum
\begin{equation}
\label{momcons}
h_{\,\,\,c}^a n^b \nabla_b J^c + D_b S^{ba} - \left(K_{\,\,\,c}^a + Kh_{\,\,\,c}^a\right)J^c
+ \left(S_{\,\,\,b}^a + E h_{\,\,\,b}^a \right) D^b[{\rm ln}N] =0\,\,\,.
\end{equation}
Of course this can be written as an evolution equation for $J^c$ by noting 
$h_{\,\,\,c}^a n^b \nabla_b J^c = h_{\,\,c}^{a} {\cal L}_{\mbox{\boldmath{$n$}}} J^c + K_{\,\,\,c}^a J^c$ .

In a similar fashion, the projection of  Eq. (\ref{consTeff}) on $n_a$, leads to the equation of conservation of energy
\begin{equation}
\label{encons}
n^b \nabla_b E + D_b J^b + 2J^b D_b[{\rm ln}N] - S^{ab} K_{ab} -  EK    =0  \,\,\,\,,
\end{equation}
which provides the evolution for $E$ since $n^b \nabla_b E= {\cal L}_{\mbox{\boldmath{$n$}}} E$.

The next step consist in using the expressions (\ref{totE})$-$(\ref{Sab}) in the equations (\ref{momcons}) and (\ref{encons}). 
It is clear that the 3+1 field equations will be required in order to simplify the equations. An extremely involved calculation that 
demands the use of the evolutions Eqs. (\ref{Pidef}), (\ref{evQ2}), (\ref{EDEfSST}) and 
(\ref{evPi1}) as well as the constraints Eqs. (\ref{CEHfSST}) and (\ref{CEMfSST}) leads finally to the desired result
\begin{equation}
\label{momconsmatt}
h_{\,\,\,c}^a n^b \nabla_b J^c_{\rm matt}  + D_b S^{ba}_{\rm matt} - \left(K_{\,\,\,c}^a + Kh_{\,\,\,c}^a\right)J^c_{\rm matt}
+ \left(S_{{\rm matt}\,\,\,b}^a + E_{\rm matt} h_{\,\,\,b}^a \right) D^b[{\rm ln}N] =0\,\,\,,
\end{equation}

\begin{equation}
\label{enconsmatt}
n^b \nabla_b E_{\rm matt} + D_b J^b_{\rm matt} + 2J^b_{\rm matt} D_b[{\rm ln}N] - S^{ab}_{\rm matt} K_{ab} -  E_{\rm matt}K    =0  \,\,\,\,.
\end{equation}

This self-consistency check ensures that the 3+1 field equations of STT in 
Sec. III are correct in their full form.

\end{document}